\documentclass[utf8]{frontiersFPHY} 
\setcitestyle{square} 
\usepackage{url,hyperref,lineno,microtype,subcaption}
\usepackage[onehalfspacing]{setspace}
\usepackage{amsmath}	
\usepackage{amssymb}	

\def\keyFont{\fontsize{8}{11}\helveticabold }
\def\firstAuthorLast{Georgakarakos {et~al.}} 
\def\Authors{Nikolaos Georgakarakos\,$^{1,2,*}$, Siegfried Eggl\,$^{4,5,6,7}$ and Ian Dobbs-Dixon\,$^{1,2,3}$}
\def\Address{$^{1}$ Division of Science, New York University Abu Dhabi, Abu Dhabi, UAE \\
$^{2}$Center for Astro, Particle and Planetary Physics (CAP$^3$), New York University Abu Dhabi,UAE\\
$^{3}$Center for Space Sciences, New York University Abu Dhabi,UAE\\
$^{4}$Department of Astronomy, University of Washington, Seattle, WA, USA\\
$^{5}$Vera C. Rubin Observatory, Tucson, Arizona, USA\\
$^{6}$Department of Aerospace Engineering, University of Chicago at Urbana-Champaign, Urbana, IL, USA\\
${7}$IMCCE, Observatoire de Paris, Paris, France}

\def\corrAuthor{Nikolaos Georgakarakos}

\def\corrEmail{ng53@nyu.edu}

\begin{document}
\onecolumn
\firstpage{1}

\title[Circumbinary habitable zones]{Circumbinary habitable zones in the presence of a giant planet} 

\author[\firstAuthorLast ]{\Authors} 
\address{} 
\correspondance{} 

\extraAuth{}

\maketitle

\begin{abstract}

\section{}
Determining habitable zones in binary star systems can be a challenging task due to the combination of perturbed planetary orbits and varying stellar irradiation conditions. The concept of ``dynamically informed habitable zones" allows us, nevertheless, to make predictions on where to look for habitable worlds in such complex environments. Dynamically informed habitable zones have been used in the past to investigate the habitability of circumstellar planets in binary systems and Earth-like analogs in systems with giant planets. Here, we extend the concept to potentially habitable worlds on circumbinary orbits. We show that habitable zone borders can be found analytically even when another giant planet is present in the system. By applying this methodology to Kepler-16, Kepler-34, Kepler-35, Kepler-38, Kepler-64, Kepler-413, Kepler-453, Kepler-1647 and Kepler-1661 we demonstrate that the presence of the known giant planets in the majority of those systems does not preclude the existence of potentially habitable worlds. Among the investigated systems Kepler-35, Kepler-38 and Kepler-64 currently seem to offer the most benign environment. In contrast, Kepler-16 and Kepler-1647 are unlikely to host habitable worlds.

\tiny
 \keyFont{ \section{Keywords:} Planet-star interactions, Celestial mechanics, Astrobiology, Circumbinary planets,
Habitable planets, Methods: analytical} 
\end{abstract}

\section{Introduction}
Over the past three decades exoplanet researchers have discovered more than
four thousands planets outside our Solar System\footnote{https://exoplanetarchive.ipac.caltech.edu/}. Improvements in detection techniques have now reached a point where finding planets of similar size to our Earth has become a reality.
Four of the planets in the Trappist-1 system \citep{2016Natur.533..221G} or TOI-700d \citep{2020AJ....160..116G,2020AJ....160..117R} are prime examples. Several planets in those systems orbit their host star inside the habitable zone. The habitable zone is the region where a terrestrial planet on a circular orbit about its host star can support liquid water on its surface \citep{1993Icar..101..108K}.  A number of planets have been found to reside in binary stars systems, some of which even orbit both stars \citep[e.g.][]{2011Sci...333.1602D,2012Natur.481..475W,2014ApJ...784...14K}. We refer to the latter as circumbinary planets.
Unresolved questions regarding the formation and dynamical evolution of such systems \cite{2019Galax...7...84M} have motivated a number of studies in recent years, particularly on whether or not such systems could host potentially habitable worlds \citep{2013ApJ...762....7K,2013ApJ...777..166H,2014MNRAS.443..260J,2014ApJ...780...14C,
2014MNRAS.437.1352F,2015ApJ...798..101C,2019AstL...45..620S,2019PASP..131l4402C,2020MNRAS.499.1506Y}. 
Some of the challenges in assessing habitability in binary star systems arises from the fact that one has to account for two sources of radiation, possibly of different spectral type. Moreover,  the distance of the planet to each star keeps changing over time in a non-trivial manner due to gravitational interactions between the planet and the two stars.

 The introduction of ``dynamically informed habitable zones" allowed \cite{2012ApJ...752...74E} to study the prospects for habitability of planets orbiting a single star in a binary star system.  Dynamically informed habitable zones for systems with a potentially habitable world on a circumbinary orbit were developed in \cite{2018haex.bookE..61E} and \cite{2020Galax...8...65E}.
In this work, we improve on previous analytic estimates for circumbinary dynamically informed habitable zones  and extend the concept to systems that are known to host an additional giant planet.  Here we refer to giant planets as bodies with masses ranging from Neptune mass up to a few Jupiter masses.

The structure of this paper is as follows: in the next section we explain the general principles behind dynamically informed habitable zones and construct the required tools to extend the concept to include a giant planet in the system. In section 3 we investigate the potential of binary star systems with known circumbinary planets observed during the Kepler mission \cite[e.g.][]{2011Sci...333.1602D,2012Natur.481..475W,2014ApJ...784...14K} to host additional potentially habitable worlds. Finally, in section 4 we summarize and discuss the work presented here.

\section{Dynamically Informed Habitable Zones}\label{sec:method}
The nearly circular orbit of the Earth around the Sun ensures that the planet receives an almost constant amount of radiation on a permanent basis. That assumption falters for a circumbinary planet, however. The second star provides an additional source of radiation, and more importantly, it is also a source of gravitational perturbations for the planetary orbit.  Even if a planet is on an initially circular orbit around the binary, the orbit will become elliptic over time \citep[see e.g.][]{2009MNRAS.392.1253G,2015ApJ...802...94G}.
As a consequence, the planet experiences time dependent irradiation. Depending on how effectively the climate on the planet can buffer changes in incoming radiation its response to radiative forcing can differ widely \citep[see e.g.][]{2017NatCo...814957P,2017ApJ...835L...1W,2019JGRE..124.3231H,2020JGRE..12506576W}. In order to capture the various responses defined by a planet's climate inertia, we make use of so-called 'dynamically informed habitable zones' (DIHZs) \citep{2018ApJ...856..155G}. DIHZs not only take the orbital evolution of the planet around the binary into account, but they can even trace habitable zone limits for different climate inertia values a planet may have. Climate inertia is defined as the time it takes climate parameters, such as the mean surface temperature, to react to radiative forcing. The faster the mean surface temperature changes, the lower the climate inertia of a planet is.

In order to account for the effect of climate inertia, we follow the general methodology as outlined in \citep{2012ApJ...752...74E} and \citep{2018ApJ...856..155G} and introduce three different DIHZs: the permanently habitable zone (PHZ), the averaged habitable zone (AHZ) and the extended habitable zone (EHZ). The PHZ is the most conservative region. For a planet to reside in the PHZ means that it stays continuously within habitable insolation limits in spite of orbit-induced variability. In other words, the PHZ is the region where a planet with essentially zero climate inertia could remain habitable on stellar evolution timescales. 
On the other end of the spectrum is the averaged habitable zone, where a planet is assumed to buffer all variations in irradiation and remains habitable as long as the insolation average stays within habitable limits.
This scenario corresponds to a planet with a very high climate inertia. 

The extended habitable zone lies between the above extremes by assuming that the planetary climate has limited buffering capabilities. The EHZ is defined as the region where the planet stays on average plus minus one standard deviation within habitable insolation limits.

In order to calculate DIHZ borders for circumbinary systems, we need to understand a) whether or not a configuration is dynamical stable, b) how the orbital evolution of the system affects the amount of radiation the planet receives and c) how the combined quantity and spectral distribution of the star light influences the climate of a potentially habitable world.\\

\subsection{Dynamical stability} 

Dynamical stability is a necessary condition for habitability of a circumbinary planet. If a planet is ejected from a system, water that may be present on its surface will ultimately freeze.  There are a number of ways to predict whether or not a potentially habitable world is on a stable orbit \citep[e.g. see][]{2008CeMDA.100..151G}.  In this work, we make use of the empirical condition developed in \cite{1999AJ....117..621H} which provides critical semi-major axis values below which planetary orbits around binary stars become unstable. The critical semi-major axis in a circumbinary system depends on the eccentricity, semi-major axis and mass ratio of the binary as can be seen from the following equation:

\begin{eqnarray}
a_{c} & = & [(1.60\pm 0.04)+(5.10\pm 0.05)e_b+(-2.22\pm 0.11)e_b^{2}+(4.12\pm 0.09)\mu\nonumber\\
& &  +(-4.27\pm 0.17)e_b\mu+(-5.09\pm 0.11)\mu^{2}+(4.61\pm 0.36)e_b^{2}\mu^{2}]a_{b},
\label{holman}
\end{eqnarray}
where $a_b$ is the semi-major axis of the binary, $e_b$ is the eccentricity of the binary and $\mu=m_2/(m_1+m_2)$; $m_1$ and $m_2$ are the masses of the two stars ($m_1 > m_2$ without loss of generality). A value of planetary semi-major axis below $a_c$
indicates that the planet will escape from the system or collide with one of the stars.

Once we confirm that a potentially habitable planet is on a stable orbit, we can proceed to investigate how much radiation it receives from the two stars. The latter depends on the orbit of the planet which evolves over time. By modeling the evolution of the orbit of the planet we can estimate the actual amount and spectral composition of the radiation the planet receives.  To this end we make use of an analytic orbit propagation technique for circumbinary planets developed in \cite{2015ApJ...802...94G}. \\ 

\subsection{The Classical Habitable Zone}

For systems consisting of a star and a terrestrial planet on a fixed circular orbit, the limits of the classical habitable zone (CHZ) read 
\begin{equation}
\label{eq:dist}
r_I=\left(\frac{L}{S_I}\right)^{\frac{1}{2}} \hspace{0.4cm}\mbox{and}\qquad r_O=\left(\frac{L}{S_O}\right)^{\frac{1}{2}},
\end{equation}
where $r$ is the distance of the planet to its host star in astronomical units, $L$ is the host star's luminosity in solar luminosities, and $S_I$ and $S_O$ 
are effective insolation values or ``spectral weights" \citep{1993Icar..101..108K}. The latter correspond to the number of solar constants that trigger a runaway greenhouse process (subscript $I$) evaporating surface oceans,
or a snowball state (subscript $O$) freezing oceans on a global scale. Spectral weights are functions of the effective temperature of the host star and, therefore, they take the specific wavelength distribution of a star's light into account. They can be calculated by the expressions below which can be found in \cite{2014ApJ...787L..29K}:

\begin{eqnarray}
\label{kop1}
S_I&=&1.107+1.332\cdot10^{-4}T_c+1.580\cdot10^{-8}T^2_c-8.308\cdot10^{-12}T^3_c-1.931\cdot10^{-15}T^4_c\\
\label{kop2}
S_O&=&0.356+6.171\cdot10^{-5}T_c+1.698\cdot10^{-9}T^2_c-3.198\cdot10^{-12}T^3_c-5.575\cdot10^{-16}T^4_c.
\end{eqnarray}
$T_c=T_{eff}/1\;K-5780$, with $T_{eff}$ being the effective temperature of the star, while the value of 5780 used in the above fit formulae corresponds to the effective temperature $T_{\odot}$ of the Sun. The coefficients in equations (\ref{kop1}) and (\ref{kop2}) refer to a planet of one Earth mass but similar functions are available for different terrestrial planet masses. In case stellar luminosities have not been observed directly, one can use stellar radii $R_*$ and effective temperatures $T_{eff}$ instead:
\begin{equation}
\label{eq:dist1}
\frac{L}{L_\odot}=\left(\frac{R_*}{R_{\odot}}\right)^2\left(\frac{T_{eff}}{T_{\odot}}\right)^4,\nonumber
\end{equation}
where $R_{\odot}$ is the radius of the Sun.

\subsection{The Permanently Habitable Zone}
\begin{figure}[h!]
\begin{center}
\includegraphics[width=0.8\textwidth]{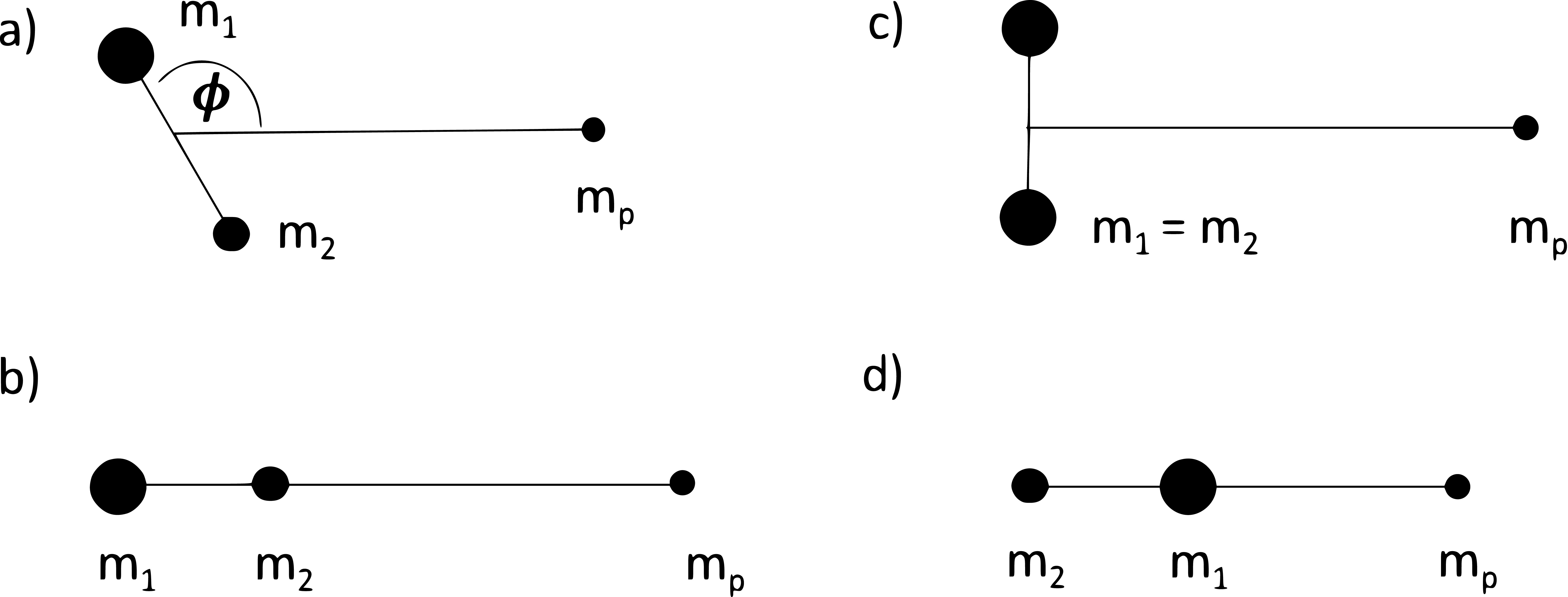}
\end{center}
\caption{A schematic representation of possible binary star - planet geometries representing irradiation minima: a), b), c) and maxima d) for various mass ratios of the binary star ($m_2/m_1$). A standard mass-luminosity relation for main sequence stars is assumed. The more massive star is $m_1$, the less massive star is $m_2$, and the planet is represented by $m_p$.}\label{fig:1}
\end{figure}

\begin{figure}[h!]
\begin{center}
\includegraphics[width=0.8\textwidth]{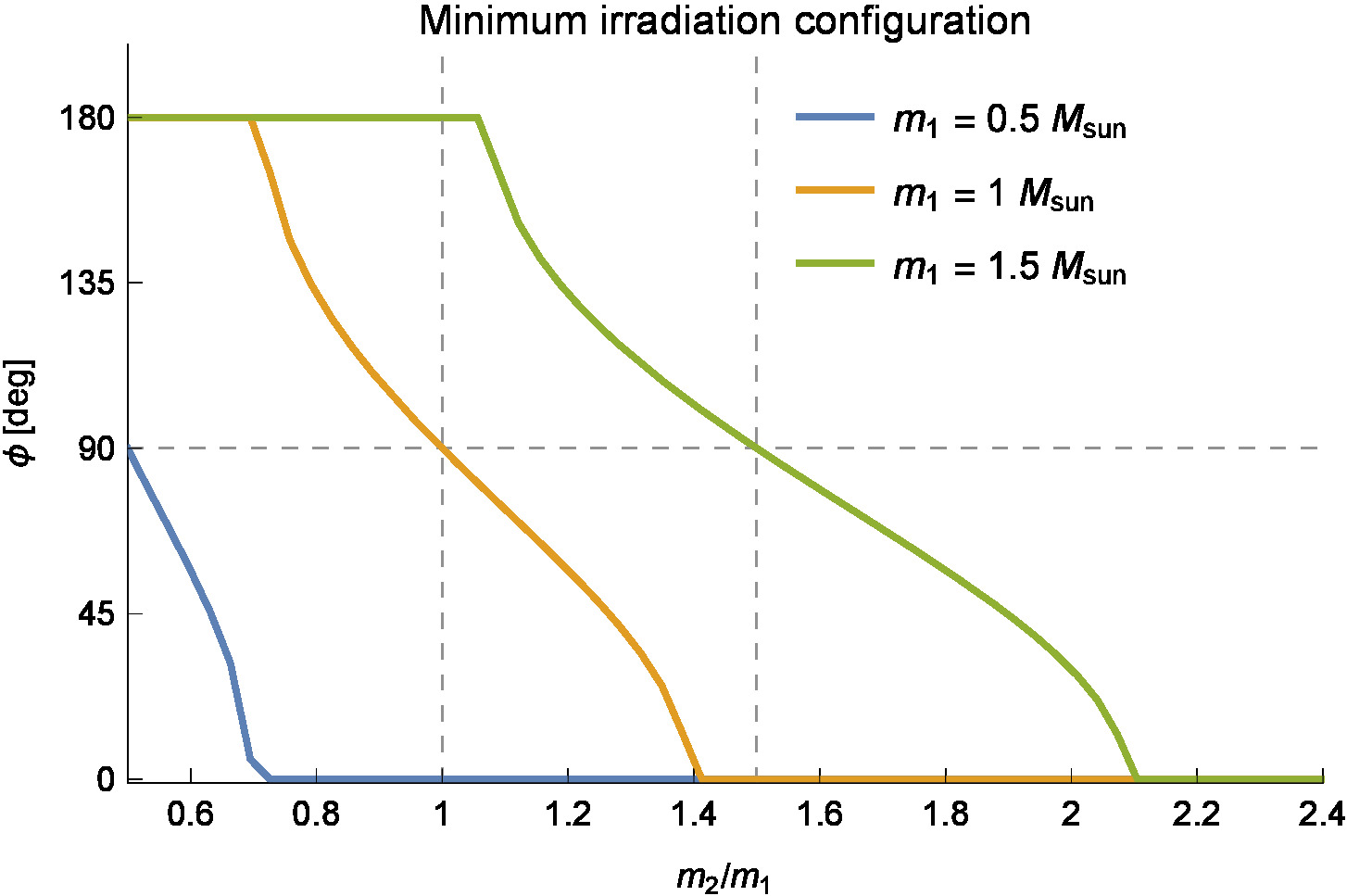}
\end{center}
\caption{Angle $\phi_{min}$ for the minimum planetary insolation configuration against binary mass ratio. We assume a planet on a circular orbit at a distance of 2.37\;au from the binary barycenter, while the binary stars are evolved on a circular orbit with a semi-major axis of 1\;au.}\label{fig:2}
\end{figure}

We will now use the above concepts to determine the permanently habitable zone.
In order to find the borders of the region wherein the planet stays always within habitable insolation limits, i.e. the PHZ, we need to find the effective insolation extrema a circumbinary planet is likely to encounter. 
In hierarchical systems of two stars and a circumbinary planet, the planetary semi-major axis remains practically constant over time. In addition, if a system is coplanar, the time evolution of the eccentricity vectors determines the geometric configuration at any given moment. Assuming furthermore that the gravitational effect of the planet on the stellar binary is negligible, maximum and minimum insolation configurations are determined through the maximum planetary eccentricity $e_p^{max}$. We use the expressions derived in \cite{2015ApJ...802...94G} that allow us to calculate to $e_p^{max}$ as a function of initial conditions and system parameters. The apocenter and pericenter distance between the planet and the barycenter of the binary star  can in turn be expressed through the maximum eccentricity as well as the semi-major axis of the planetary orbit $a_p$, i.e. $Q_p=a_p(1+e_p^{max})$ and $q_p=a_p(1-e_p^{max})$.
Figure \ref{fig:1} is a schematic representation of our system, meant to help the reader visualize configurations that lead to insolation extrema. The planet receives maximum insolation when it is at pericenter with respect to the binary barycenter and closer to the brighter star, i.e. when the angle $\phi$ between the line that connects the stars and the line that connects the planet to the binary barycenter is $\phi=0^\circ$. Thus, the three bodies are aligned and the mathematical expression that determines the inner edge of the PHZ is
\begin{equation}
PHZ_I: \frac{L_1}{S_{1,I}(q_p-\mu Q_b)^2}+\frac{L_2}{S_{2,I}[q_p+(1-\mu)Q_b]^2}\leq 1,\label{eq:5}
\end{equation}
where the subscript $p$ refers to the planet and the subscript $b$ refers to the binary star orbit.
Note that we have normalized the stellar luminosities with their respective spectral weights in order to account for the effect of each star's individual spectral energy distribution on the planetary climate.
The minimum insolation configuration is not as straight forward to determine. If one star is substantially more massive and luminous than the other a minimum can be reached when the brighter star is farthest from the planet and the planet is at apocenter with respect to the barycenter of the binary star. This is the case when $\phi=180^\circ$ which corresponds to the following condition for the outer edge of the PHZ
\begin{equation}
PHZ_O: \frac{L_1}{S_{1,O}(Q_p+\mu Q_b)^2}+\frac{L_2}{S_{2,O}[Q_p-(1-\mu)Q_b]^2}\geq 1,
\end{equation}
If we have two stars of equal mass, then 
the minimum radiation configuration is reached when $\phi=90^\circ$.
Hence, the minimum insolation condition in this case is
\begin{equation}
PHZ_O: \frac{1}{Q_p^2+0.5 Q_b^2}\left(\frac{L_1}{S_{1,O}}+\frac{L_2}{S_{2,O}}\right)\geq 1.
\end{equation}
When the mass ratio of the binary is close to - but not exactly - equal to one (with $m_1 > m_2$) and the planet is located at a suitable distance, the minimum insolation geometry occurs between $\phi=90^\circ$ and $\phi=180^\circ$. Figure \ref{fig:2} illustrates this behavior for a variety of binary mass ratios. We note, however that the difference in insolation received by the planet between said configurations is minute. Hence, we limit our approach to comparing the perpendicular and straight line configurations and use the one that provides the smallest value. Consequently the minimum insolation condition for the outer border of the PHZ reads
\begin{eqnarray}
PHZ_O: & & min\left\{\frac{L_1}{S_{1,O}(Q_p+\mu Q_b)^2}+\frac{L_2}{S_{2,O}[Q_p-(1-\mu)Q_b]^2},\right.\nonumber\\ 
& &\left. \frac{L_1}{S_{1,O}(Q_p^2+\mu^2 Q_b^2)}+\frac{L_2}{S_{2,O}[Q_p^2+(1-\mu)^2Q_b^2]}\right\}\geq 1.\label{eq:8}
\end{eqnarray}
One can find the numeric values for the borders of the PHZ by solving equations (\ref{eq:5}) through (\ref{eq:8}) for $a_p$. 

We would like to point out here that for all the above insolation extrema configurations we have assumed that the stars are point masses. In reality, however, the stars have finite sizes.
Depending on the distance between the two stars and on the distance between the planet and the binary, it is possible that when the three bodies are aligned the planet may receive reduced insolation due to the eclipsed star. In such a scenario, the minimum insolation configuration (panel (c) in Figure 1) will still be valid.  On the other hand, the maximum insolation configuration (panel (d) of Figure 1) would provide a smaller insolation value. \cite{2020JGRE..12506576W} have shown, however, that eclipses have little effect on the overall stability of the climate. But even if the insolation change is considerable, as soon as the three bodies get out of alignment, the planet will then receive near maximum insolation. Hence, equation (5) constitutes a reasonable approximation for our purposes.


%
\vspace{0.4cm}
\subsection{The Averaged Habitable Zone}
The averaged habitable zone is the region around a binary star where a planet remains habitable inspite of variations in irradiation. That is, as long as the insolation average is compatible with habitable limits a planet with a high climate inertia can remain potentially habitable. The averaged over time radiation that a planet receives when orbiting a single star is
\begin{equation}
\label{eq:sav}
\langle S \rangle=\frac{1}{P}\int_0^{P}\frac{L}{r_{p}^2(t)}dt=\frac{n_p}{2\pi}\int_0^{2\pi}\frac{Ldf_p}{n_pa_p^2\sqrt{1-e_p^2}}=
\frac{L}{a_p^2\sqrt{1-e_p^2}},
\end{equation}
where $L$ is the stellar luminosity, $P$ the period of the planetary motion, $n_p$ is the mean motion of the planet, $f_p$ is its true anomaly and $r_p$ the distance between the source of radiation and the planet; we made use of the well known relation $r_p^2df_p/dt=n_pa_p^2\sqrt{1-e_p^2}$.  We can now extend this simple relation to circumbinary orbits. Assuming that the distance between the two stars is small compared to the distance between the planet and the binary star barycenter,  i.e. the orbital period of the binary pair is much smaller than that of the planet, we can approximate the average over time insolation around the binary by placing the two stars at their barycenter and make use of equation (\ref{eq:sav}). In other words we approximate the three body system as a two-body problem with a central ``hybrid star" that has the combined mass, luminosity and spectral energy distribution of the stellar binary. This leads to 
\begin{equation}
\label{eq:savbin}
\langle S \rangle=\frac{L_1+L_2}{a_p^2\sqrt{1-\langle e_p^2 \rangle}},
\end{equation}
where $\langle e_p^2 \rangle$ is the averaged square planetary eccentricity over time and initial angles as given in \cite{2015ApJ...802...94G}. Under those assumptions the borders of the AHZ are defined through the following inequalities
\begin{equation}
AHZ_I: \frac{L_1/S_{1,I}+L_2/S_{2,I}}{a_p^2\sqrt{1-\langle e_p^2 \rangle}}\leq 1 \label{eq:AHZI}
\end{equation}
and
\begin{equation}
AHZ_O: \frac{L_1/S_{1,O}+L_2/S_{2,O}}{a_p^2\sqrt{1-\langle e_p^2 \rangle}} \geq 1.\label{eq:AHZO}
\end{equation}
Note that $\langle e_p^2 \rangle$ generally depends on $a_p$ in a non-trivial way. The full expression for $\langle e_p^2 \rangle$ along with the expression for $e_p^{max}$ is provided in the Appendix.
Once more, solving equations (\ref{eq:AHZI}) and (\ref{eq:AHZO}) for $a_p$ yields the numeric values for the borders of the AHZ. 
\vspace{0.4cm}
\subsection{The Extended Habitable Zone}
The definition of the extended habitable zone in section \ref{sec:method} translates into the following equations:
\begin{equation}
\langle S \rangle +\sigma=S_I \hspace{0.5cm} \mbox{and} \hspace{0.5cm} \langle S \rangle-\sigma=S_O.
\end{equation}

The standard deviation $\sigma$ can be found via the insolation variance
\begin{equation}
\sigma^2=\langle S^2 \rangle -\langle S \rangle^2.
\end{equation}
We already have an expression for $\langle S \rangle$ from equation (\ref{eq:savbin}), but we are yet to find $\langle S^2 \rangle$. For a planet around a single star, $\langle S^2 \rangle$ is
\begin{equation}
\label{eq:sav3}
\langle S^2 \rangle=\frac{1}{P}\int_0^{P}\frac{L^2}{r_{p}^4(t)}dt=\frac{n_p}{2\pi}\int_0^{2\pi}\frac{(1+e_p\cos{f_p})^2}{a^2_p(1-e_p^2)^2}\frac{L^2df_p}{n_pa_p^2\sqrt{1-e_p^2}}=\frac{L^2(1+e_p^2/2)}{a^4_p(1-e_p^2)^{5/2}}.
\end{equation}
Using the same approach that lead to equation (\ref{eq:savbin}), namely combining the stellar binary into a ``hybrid star" we can construct a closed analytic expression for $\langle S^2 \rangle$ of a circumbinary planet, namely
\begin{equation}
\label{eq:sav4}
\langle S^2 \rangle=\frac{(L_1+L_2)^2(1+\langle e_p^2 \rangle /2)}{a^4_p(1-\langle e_p^2 \rangle)^{5/2}}.
\end{equation}

Combining equations (\ref{eq:savbin}) and (\ref{eq:sav4}) and normalizing the individual stellar contributions with the respective spectral weights $X_i \in \{S_{i,I},S_{i,O}\}$, where $i$ is the index of the respective star, yields:

\begin{equation}
\sigma^2_X=\left(\frac{L_1}{X_1}+\frac{L_2}{X_2}\right)^2\frac{1}{a_p^4(1-\langle e_p^2 \rangle)}\left[\frac{1+\langle e_p^2 \rangle/2}{(1-\langle e_p^2\rangle)^{3/2}}-1\right]. 
\end{equation}

The inner border of the EHZ is then defined through
\begin{equation}
EHZ_I: \left(\frac{L_1}{S_{1,I}}+\frac{L_2}{S_{2,I}}\right)\frac{1}{a_p^2\sqrt{1-\langle e_p^2 \rangle}}\left\{1+\sqrt{\left[\frac{1+\langle e_p^2 \rangle/2}{(1-\langle e_p^2\rangle)^{3/2}}-1\right]}\right\}\leq 1, 
\end{equation}
while the outer border is defined via
\begin{equation}
EHZ_O: \left(\frac{L_1}{S_{1,O}}+\frac{L_2}{S_{2,O}}\right)\frac{1}{a_p^2\sqrt{1-\langle e_p^2 \rangle}}\left\{1-\sqrt{\left[\frac{1+\langle e_p^2 \rangle/2}{(1-\langle e_p^2\rangle)^{3/2}}-1\right]}\right\}\geq 1
\end{equation}
Numerical values for the inner and outer EHZ borders can be found via solving the above equations 
for $a_p$.

\section{Application to Kepler Systems with known Circumbinary Planets}
\label{sec:application}
In order to demonstrate the merit of dynamically informed habitable zones, we apply our method to the Kepler circumbinary planets. For that purpose we select Kepler-16 \citep{2011Sci...333.1602D}, Kepler-34 and Kepler-35 \citep{2012Natur.481..475W}, Kepler-38 \citep{2012ApJ...758...87O}, Kepler-64 \citep{2013ApJ...768..127S}, Kepler-413 \citep{2014ApJ...784...14K}, Kepler-453 \citep{2015ApJ...809...26W}, Kepler-1647 \citep{2016ApJ...827...86K} and Kepler-1661 \citep{2020AJ....159...94S}.  Kepler-47 \citep{2019AJ....157..174O} has three planets and is, therefore, beyond the limits of our model. The masses of Kepler-38b, Kepler-64b, Kepler-453b are not well defined. For those cases, we have decided to use the upper limit provided in the relevant publications. The parameters necessary for our calculations of the selected systems can be found in Tables \ref{tab1} (binary star) and \ref{tab2} (planet).

\begin{table}
  \caption{Physical parameters and orbital elements for the Kepler-16(AB), Kepler-34(AB), Kepler-35(AB),  Kepler-38(AB), Kepler-64(AB), Kepler-413(AB), Kepler-453(AB), Kepler-1647(AB) and Kepler-1661(AB) stellar binaries.} 
 \label{tab1}
\begin{center}	
\begin{tabular}{l l l l l l l l l}
\hline\\
    System & $ M_1 (M_{\odot})$ & $ M_2 (M_{\odot})$ & $ R_1 (R_{\odot})$ & $ R_2 (R_{\odot})$ & $ T_{eff1} (K)$ & $T_{eff2}(K)$ & $a_b (au)$ & $e_b$ \\
      \\
    \hline\\
    Kepler-16 & 0.6897 & 0.20255 & 0.6489 & 0.22623 & 4450.0 & 3311.0 & 0.22431 & 0.15944 \\
    Kepler-34 & 1.0479 & 1.0208 & 1.1618 & 1.0927 & 5913.0 & 5867.0 & 0.22882 & 0.52087 \\
    Kepler-35 & 0.8876 & 0.8094 & 1.0284 & 0.7861 & 5606.0 & 5202.0 & 0.17617 & 0.1421 \\
    Kepler-38 & 0.949 & 0.249 & 1.757 & 0.2724 & 5640.0 & 3325.0 & 0.1469 & 0.1032 \\
    Kepler-64 & 1.528 & 0.378 & 1.734 & 0.408 & 6407.0 & 3561.0 & 0.1744 & 0.2117 \\
    Kepler-413 & 0.820 & 0.5423 & 0.7761 & 0.484 & 4700.0 & 3463.0 & 0.10148 & 0.0365\\
    Kepler-453 & 0.944 & 0.1951 & 0.833 & 0.2150 & 5527.0 & 3226.0 & 0.18539 & 0.0524 \\
    Kepler-1647 & 1.210 & 0.975 & 1.7903 & 0.9663 & 6210.0 & 5770.0 & 0.1276 & 0.1593 \\
    Kepler-1661 & 0.841 & 0.262 & 0.762 & 0.276 & 5100.0 & 3585.0 & 0.187 & 0.112 \\  \\
       \hline
  \end{tabular}
\end{center}
\end{table}

\begin{table}
  \caption{Mass, semi-major axis and eccentricity for Kepler-16b, Kepler-34b, Kepler-35b,  
Kepler-38b, Kepler-64b, Kepler-413b, Kepler-453b, Kepler-1647b and Kepler-1661b.} 
 \label{tab2}
\begin{center}	
\begin{tabular}{l l l l}\hline\\
    System & $m_p(M_J)$ & $a_p$ (au) & $e_p$ \\ \\
    \hline\\
    Kepler-16 & 0.333 & 0.7048 & 0.00685 \\
    Kepler-34 & 0.22 & 1.0896 & 0.182 \\
    Kepler-35 & 0.127 & 0.60345 & 0.042 \\
    Kepler-38 & $<0.384 \hspace{0.1cm}(95\%$ conf.) & 0.4644 & $<0.032 \hspace{0.1cm}(95\%$ conf.) \\
    Kepler-64 & $<0.531 \hspace{0.1cm}(99.7\%$ conf.)& 0.634 & 0.0539 \\
    Kepler-413 & 0.21 & 0.3553 & 0.1181 \\
    Kepler-453 & $<0.050$ & 0.7903 & 0.0359 \\
    Kepler-1647 & 1.52 & 2.7205 & 0.0581 \\
    Kepler-1661 & 0.053 & 0.633 & 0.057 \\

\\
       \hline
  \end{tabular}
\end{center}
\end{table}

\begin{table}
  \caption{Habitable zone limits for the Kepler-16, Kepler-34, Kepler-35, Kepler-38, Kepler-64, Kepler-413,
Kepler-453, Kepler-1647 and Kepler-1661. RHZ is the radiative habitable zone described in \cite{2014ApJ...780...14C,2015ApJ...798..101C,2019ApJ...873..113W}, while the H$\&$K column presents the results of \cite{2013ApJ...777..166H}. Finally, the column CHZ-K13 provides values for our classical zone using \cite{2013ApJ...770...82K} as
that specific version of the climate model is also used by \cite{2013ApJ...777..166H}.}
 \label{tab3}
\begin{center}	
\begin{tabular}{l l l l l l l l}\hline\\
    System & CHZ (au)& PHZ (au) & EHZ (au) & AHZ  (au) & RHZ (au) & H$\&$K (au)& CHZ-K13 (au)\\ \\
    \hline\\
    Kepler-16 & 0.40 - 0.74 & \hspace{0.85cm}- & \hspace{0.85cm}- & \hspace{0.85cm} - &0.46 - 0.70 & 0.40 - 0.76 & 0.41 - 0.74 \\
    Kepler-34 & 1.56 - 2.75 & 2.10 - 2.25 & 1.79 - 2.47 & 1.60 - 2.76 & 1.60 - 2.74  & 1.51 - 2.85 & 1.62 - 2.77 \\
    Kepler-35 & 1.12 - 1.99 & 1.23 - 1.90 & 1.15 - 1.96 & 1.12 - 1.99 & 1.16 - 1.97 & 1.09 - 2.10 & 1.16 - 2.01\\
    Kepler-38 & 1.61 - 2.84 & 1.68 - 2.77 & 1.62 - 2.82 & 1.61 - 2.84 & 1.64 - 2.81 & 1.63 - 2.82 & 1.66 - 2.86\\
    Kepler-64 & 1.96 - 3.41 & 2.08 - 3.28 & 2.00 - 3.36 & 1.96 - 3.41 & 2.00 - 3.37 & 2. 00 - 3.40 & 2.02 - 3.44 \\
    Kepler-413 & 0.55 - 1.01 & 0.69 - 0.87 & 0.60 - 0.95 & 0.55 - 1.01 & 0.58 - 0.98 & \hspace{0.85cm}- & \hspace{0.85cm}- \\
    Kepler-453 & 0.74 - 1.31 & 1.00 - 1.20 & 1.21 - 1.27 & 1.27 - 1.31 & 0.77 -1.28 & \hspace{0.85cm}- &\hspace{0.85cm}- \\
    Kepler-1647 & 2.12 - 3.71 & \hspace{0.85cm}- & \hspace{0.85cm}- & \hspace{0.85cm}- & 2.16 - 3.67 & \hspace{0.85cm}-& \hspace{0.85cm}-\\
    Kepler-1661 & 0.60 - 1.08 & 0.82 - 0.64 & 0.94 - 1.03 & 1.03 - 1.08  & 0.64 - 1.04 &\hspace{0.85cm}-& \hspace{0.85cm}-\\
\\
       \hline
  \end{tabular}
\end{center}
\end{table}

First, we calculate all dynamically informed habitable zones assuming no giant planets are present in the systems. That provides us with a first idea of the location of the habitable zones and how the presence of the second star affects the location and extent of the various habitable zones. Then we include the existing giant planet in our model and examine its effect on the habitability of an additional hypothetical terrestrial planet. In both stages, we allowed the eccentricities of the binary and the existing planet to vary. That way we get a better picture of the effect of orbital eccentricity, an important quantity that regulates distances between bodies, on the extent of habitable zones in the system.   
In order to simplify the complex dynamics in the presence of the giant planet, we acknowledge the double hierarchical structure of the problem that allows us to consider the binary as one massive body located at the barycenter.  The two stars at their barycenter are considered as one body of mass $m_b=m_1+m_2$, therefore reducing the four-body problem to a three-body one.  To describe the dynamical evolution of such a system we made use of the relevant equations in \cite{2015ApJ...802...94G} when $a_{gp} < a_p$  and \cite{2012ApJ...752...74E,2003MNRAS.345..340G,2005MNRAS.362..748G} when $a_{gp} > a_p$, where $a_{gp}$ is the semi-major axis of the orbit of the giant planet.  The stability of that particular kind of triple system was assessed using the criterion developed in \cite{2015ApJ...808..120P}. The corresponding empirical formula is based on numerical simulations of a star and two planets.  The planet-star mass ratios investigated in \cite{2015ApJ...802...94G} ranged from $10^{-4}$ to $10^{-2}$, the ratio of the planetary semi-major axes was in the interval $[3,10]$, while the planetary eccentricities took values in the interval $[0,0.9]$.

A terrestrial planet in a star-planet-planet system is stable against either ejections or collisions with the central object when
\begin{equation}
\label{petro1}
\frac{a_p}{a_{gp}(1+e_{gp})} > 2.4\left(\frac{m_{gp}}{m_1+m_2}\right)^{1/3}\left(\frac{a_p}{a_{gp}}\right)^{1/2}.
\end{equation}
Here, $m_{gp}$ is the mass of the giant planet and $e_{gp}$ its orbital eccentricity.
The above expression is valid when the giant planet is closer to the binary than the terrestrial planet. If the giant planet orbits externally to the terrestrial planet, then the criterion becomes
\begin{equation}
\label{petro2}
\frac{a_{gp}(1-e_{gp})}{a_p} > 2.4\left(\frac{m_{gp}}{m_1+m_2}\right)^{1/3}\left(\frac{a_{gp}}{a_p}\right)^{1/2}.
\end{equation}
Strictly speaking, the planetary systems investigated here do not fall in the planet-star mass ratios investigated in \cite{2015ApJ...802...94G}.  The planet to planet mass ratio, however, remains within those limits which is ultimately more important for the validity of the stability criterion at hand.  This  hypothesis can be supported by the reasonable agreement between the stability estimates given by equations (\ref{petro1}) (\ref{petro2}) and those in \cite{2013NewA...23...41G}, where hierarchical triple systems with a wide range of masses and on initially circular orbits where integrated numerically and their stability limit was determined. 

\begin{figure}[h!]
\begin{center}
\includegraphics[width=8cm]{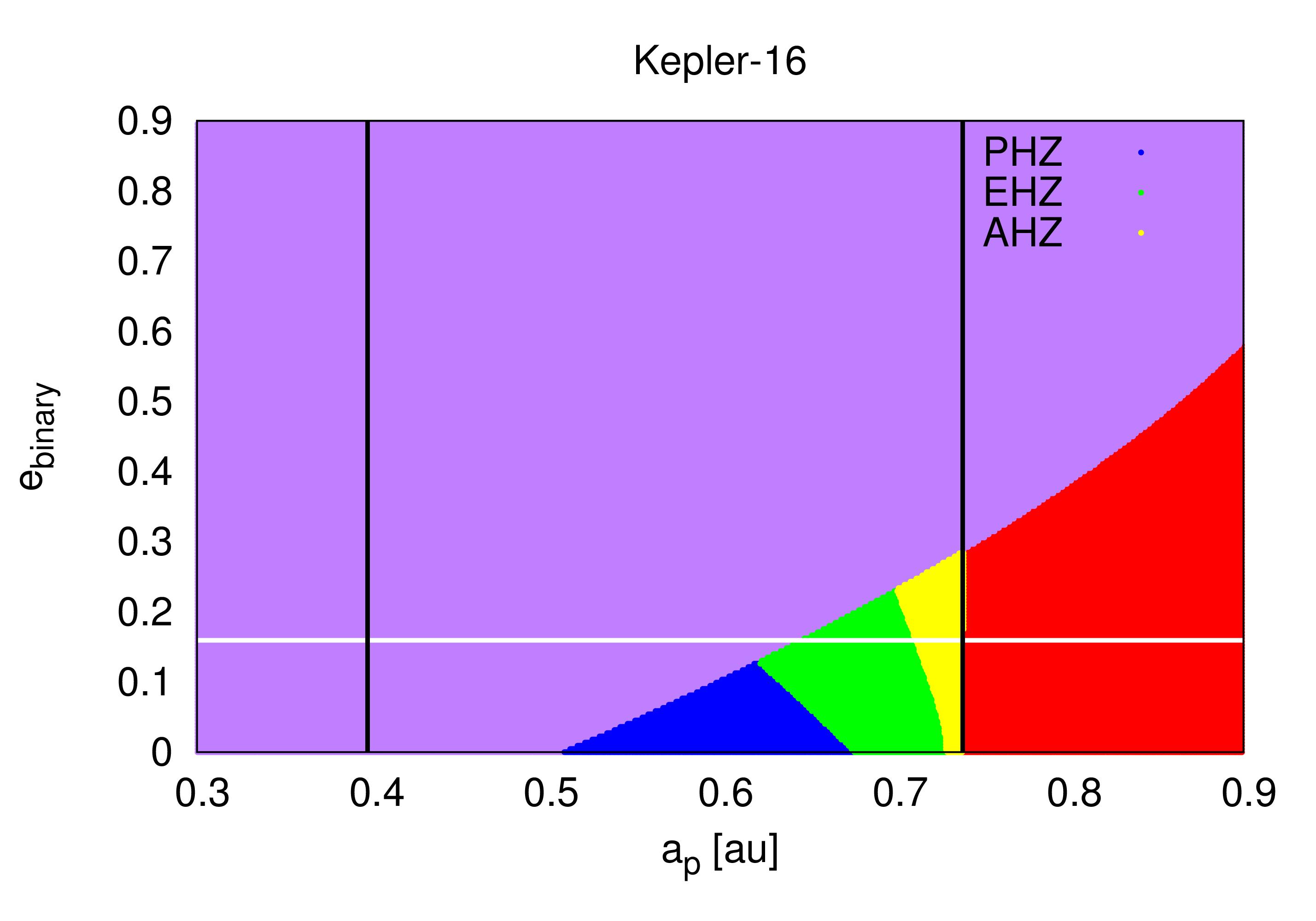}
\includegraphics[width=8cm]{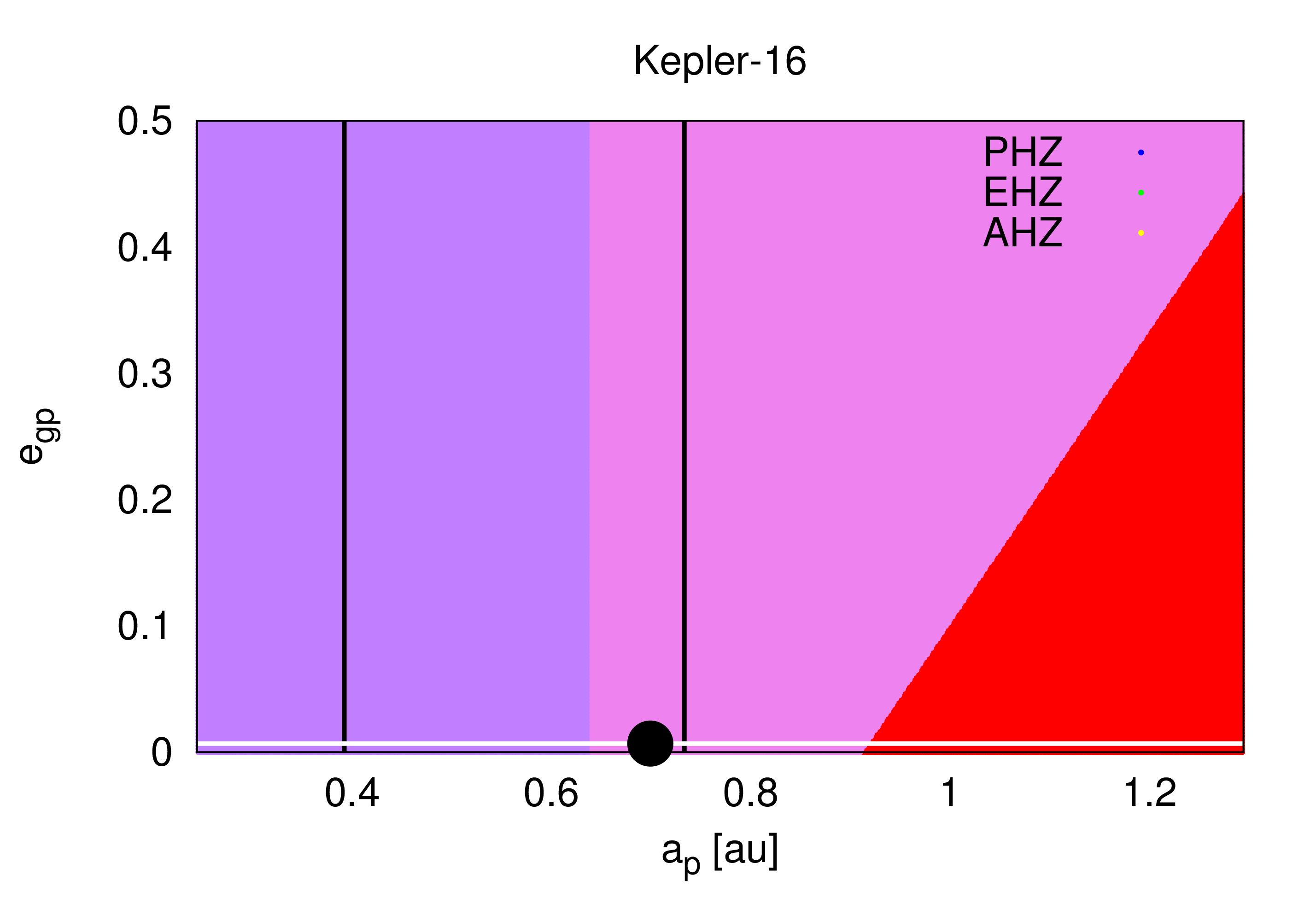}
\includegraphics[width=8cm]{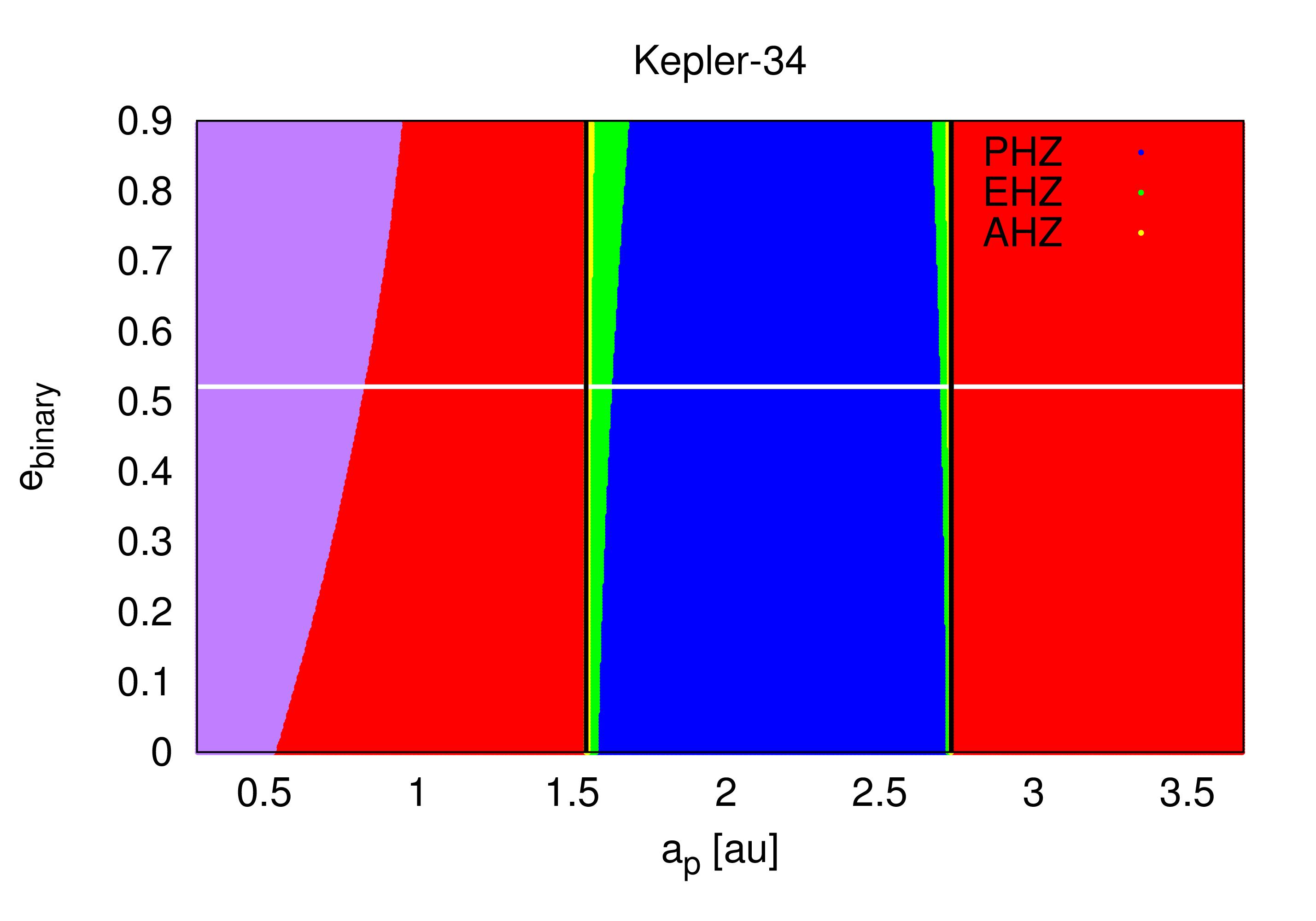}
\includegraphics[width=8cm]{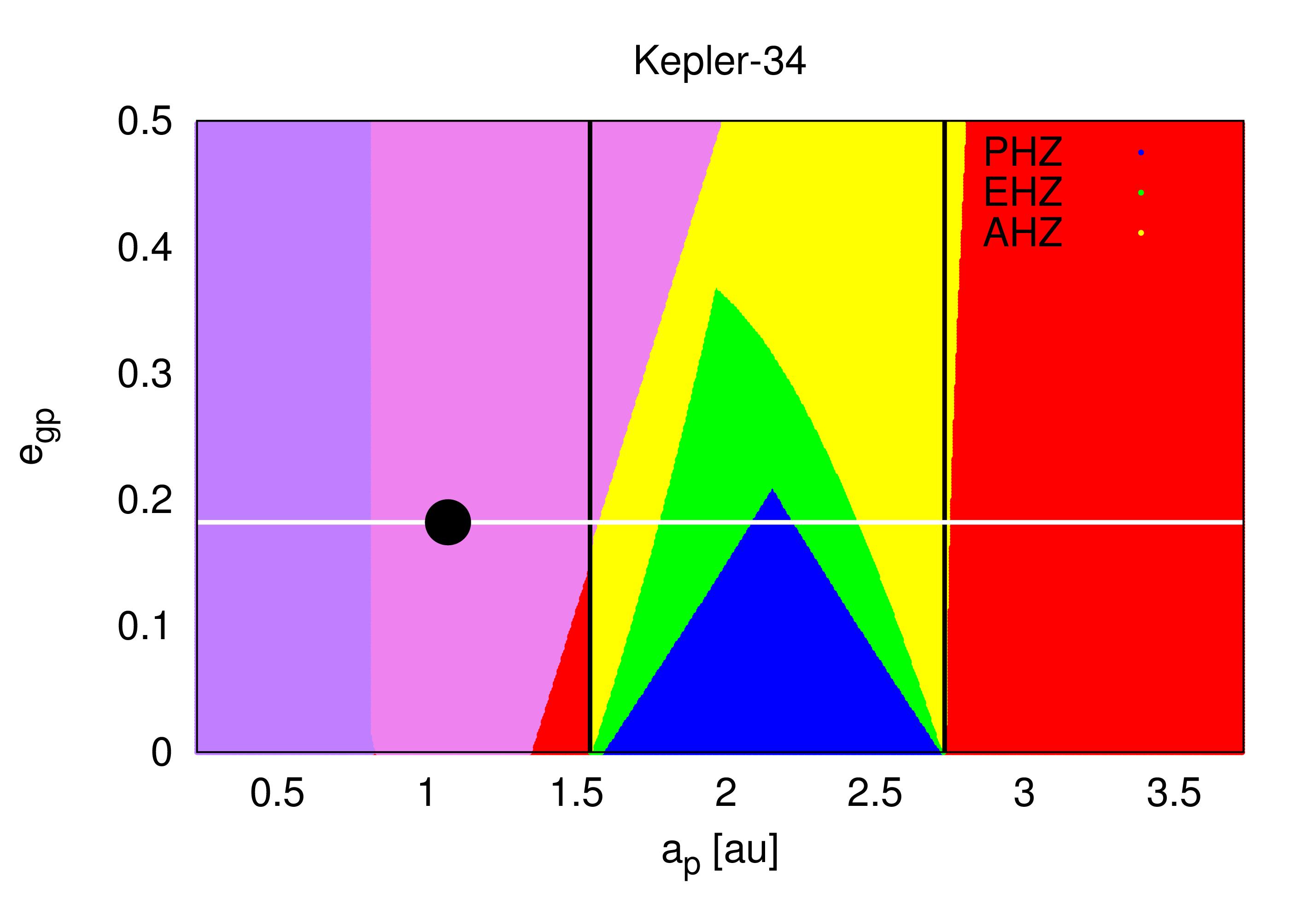}
\includegraphics[width=8cm]{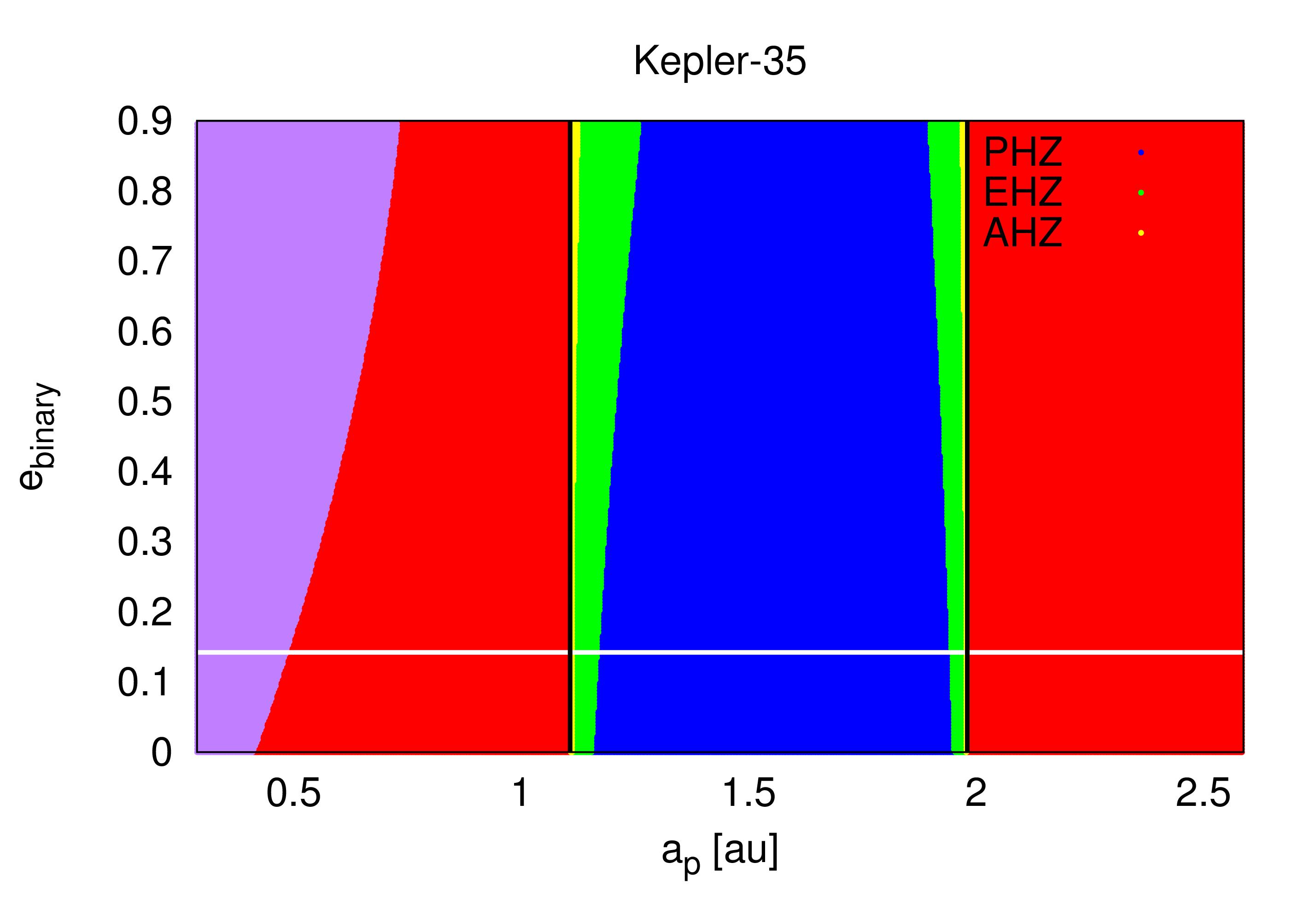}
\includegraphics[width=8cm]{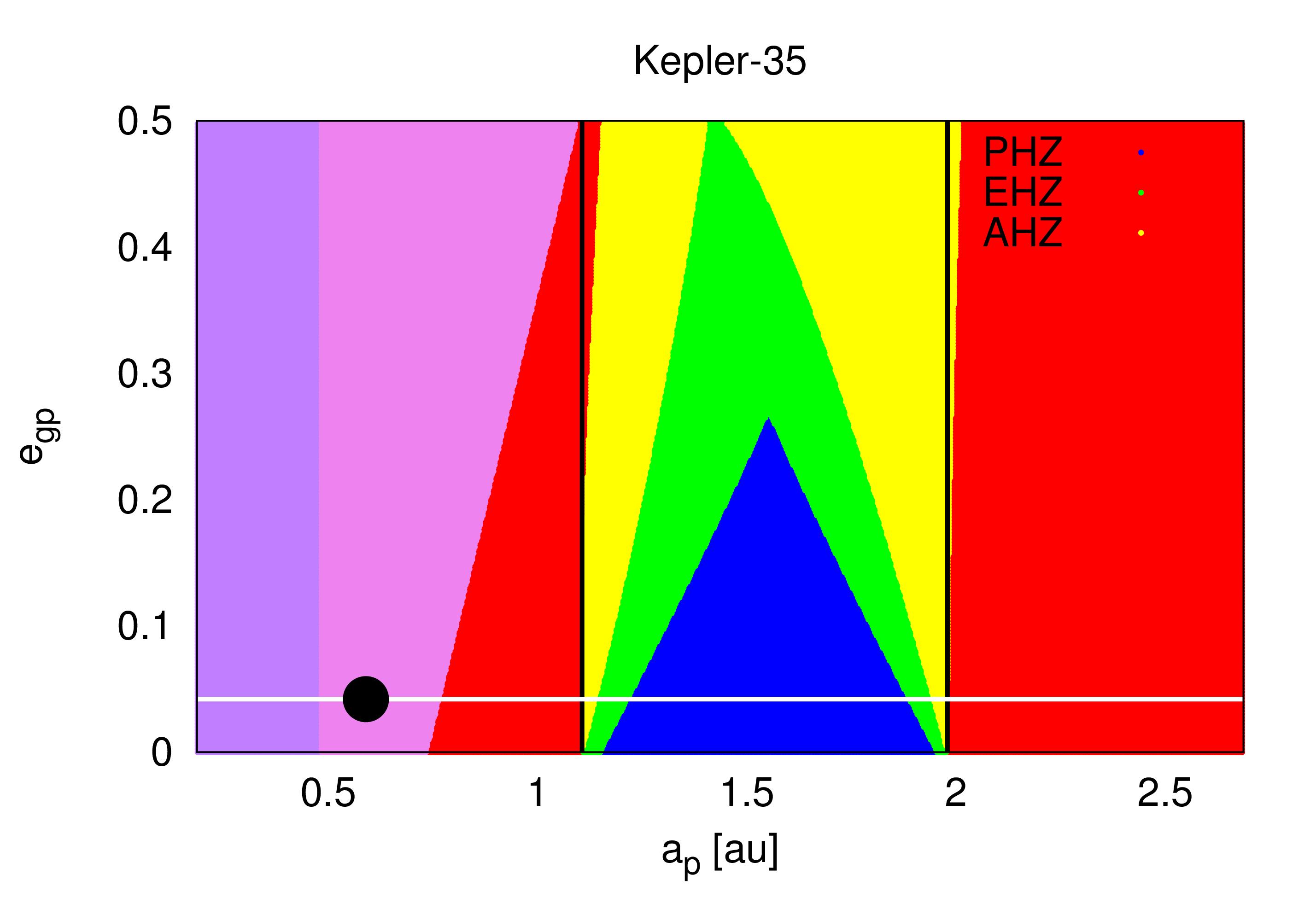}
\end{center}
\caption{Dynamically informed habitable zones for the Kepler-16, Kepler-34 and Kepler-35 systems. Plots in the left column show the different types of habitable zones without the presence of the known giant planets. The right column includes the influence of the known giant planets. Red colored regions correspond to uninhabitable areas, blue, green, yellow and purple colors denote the PHZ, the EHZ, the AHZ, and unstable areas according to \cite{1999AJ....117..621H} stability criterion, respectively. Violet colored areas mark regions of dynamical instability caused by the giant planet in the system (\cite{2015ApJ...808..120P} dynamical stability criterion). The vertical black lines denote the classical habitable zone limits, while the horizontal white line in the left column plots marks the current eccentricity of the binary star orbit. In the right column graphs, the white line marks the current eccentricity of the giant planet orbit. Finally, the black dot in the right column plots shows the position of the giant planet in the presented parameter space.}\label{fig:3}
\end{figure}

\begin{figure}[h!]
\begin{center}
\includegraphics[width=8cm]{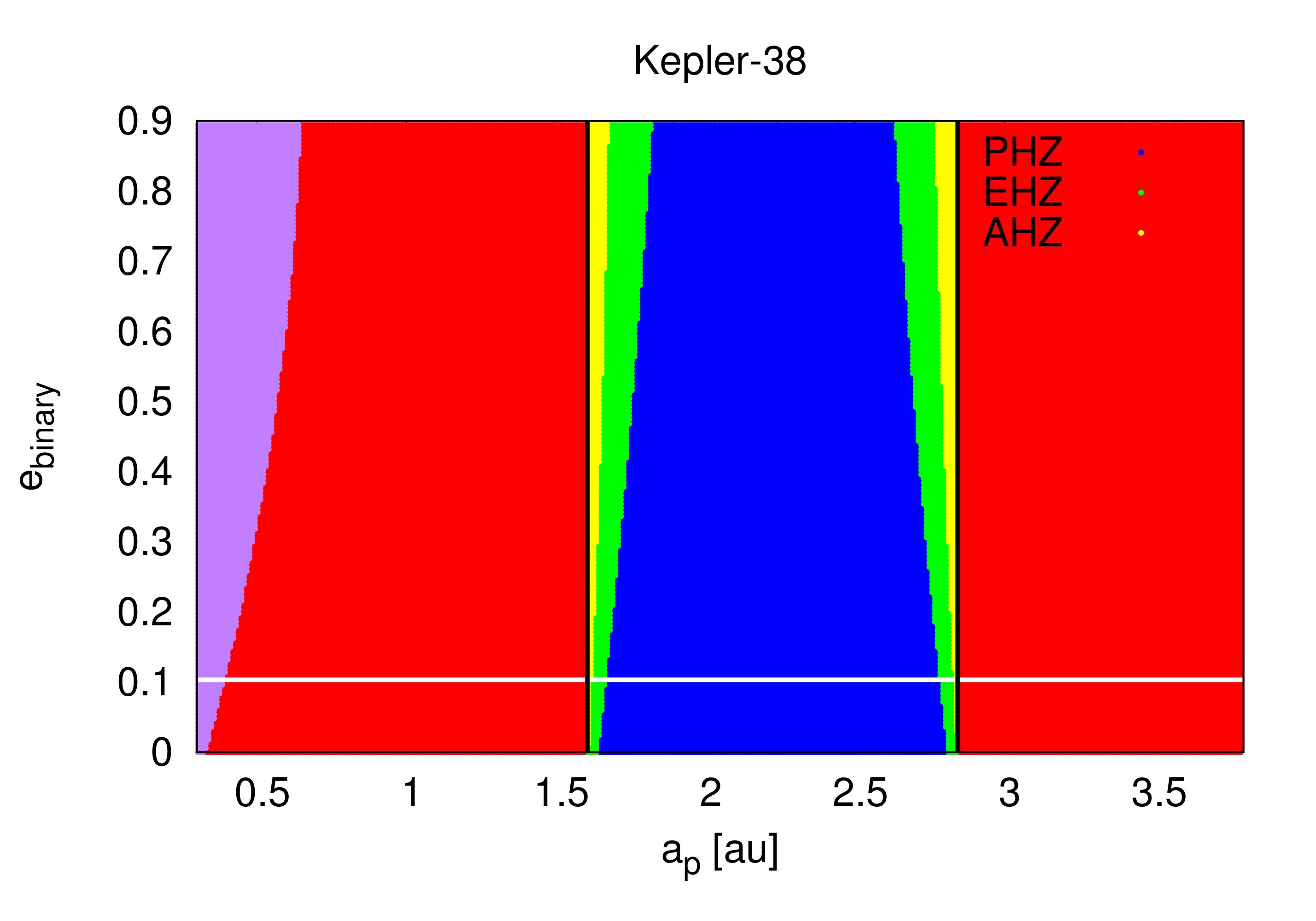}
\includegraphics[width=8cm]{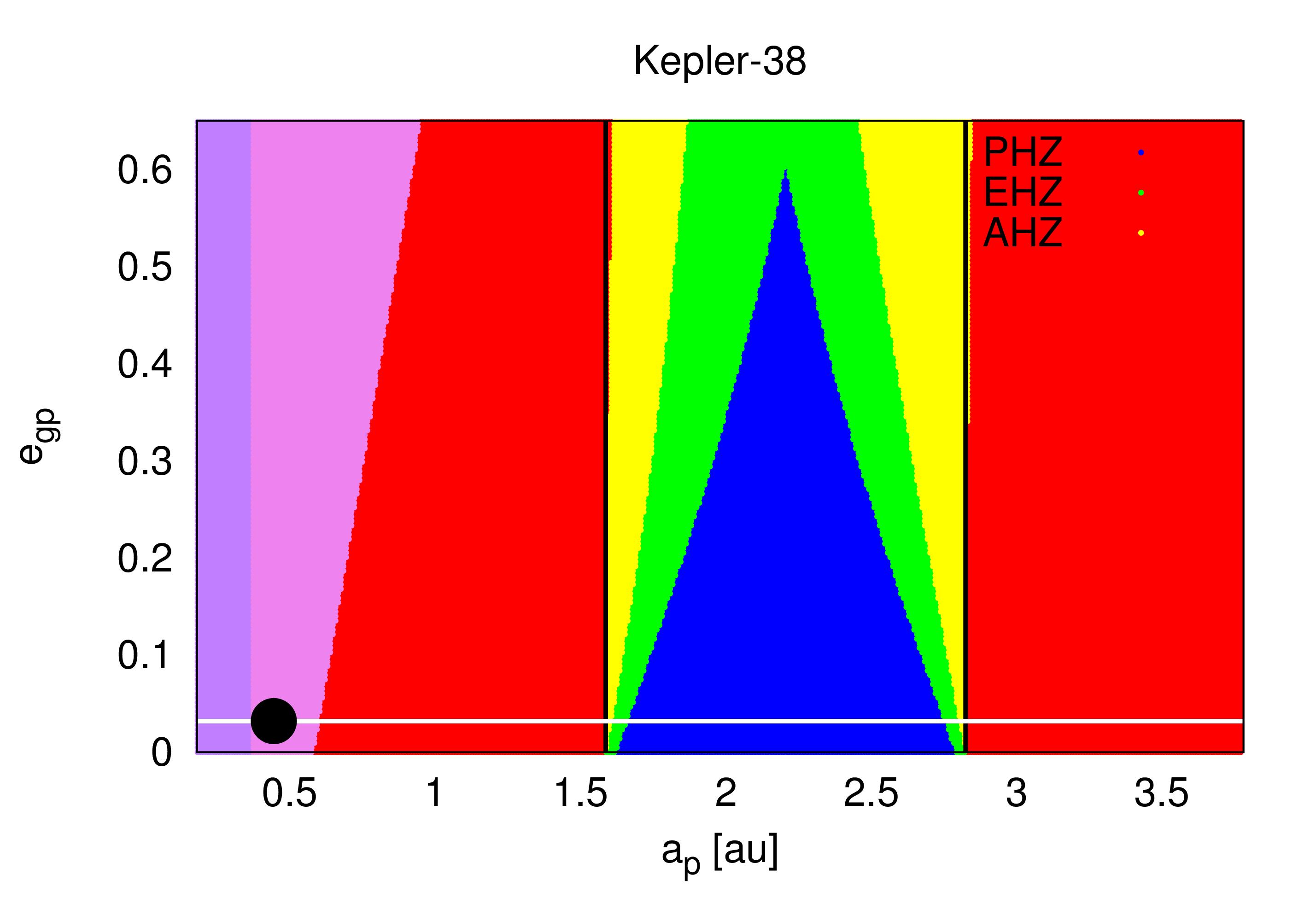}
\includegraphics[width=8cm]{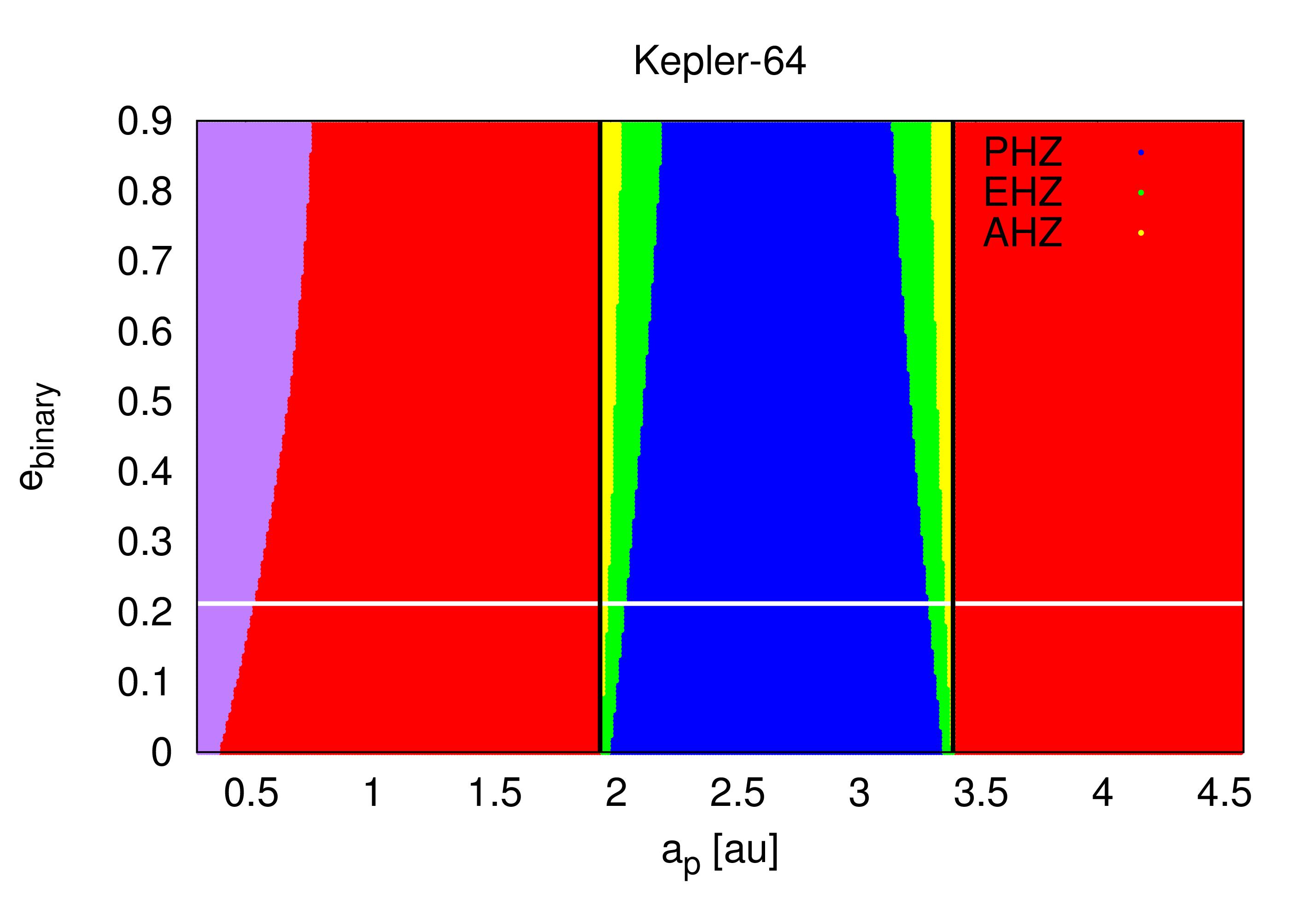}
\includegraphics[width=8cm]{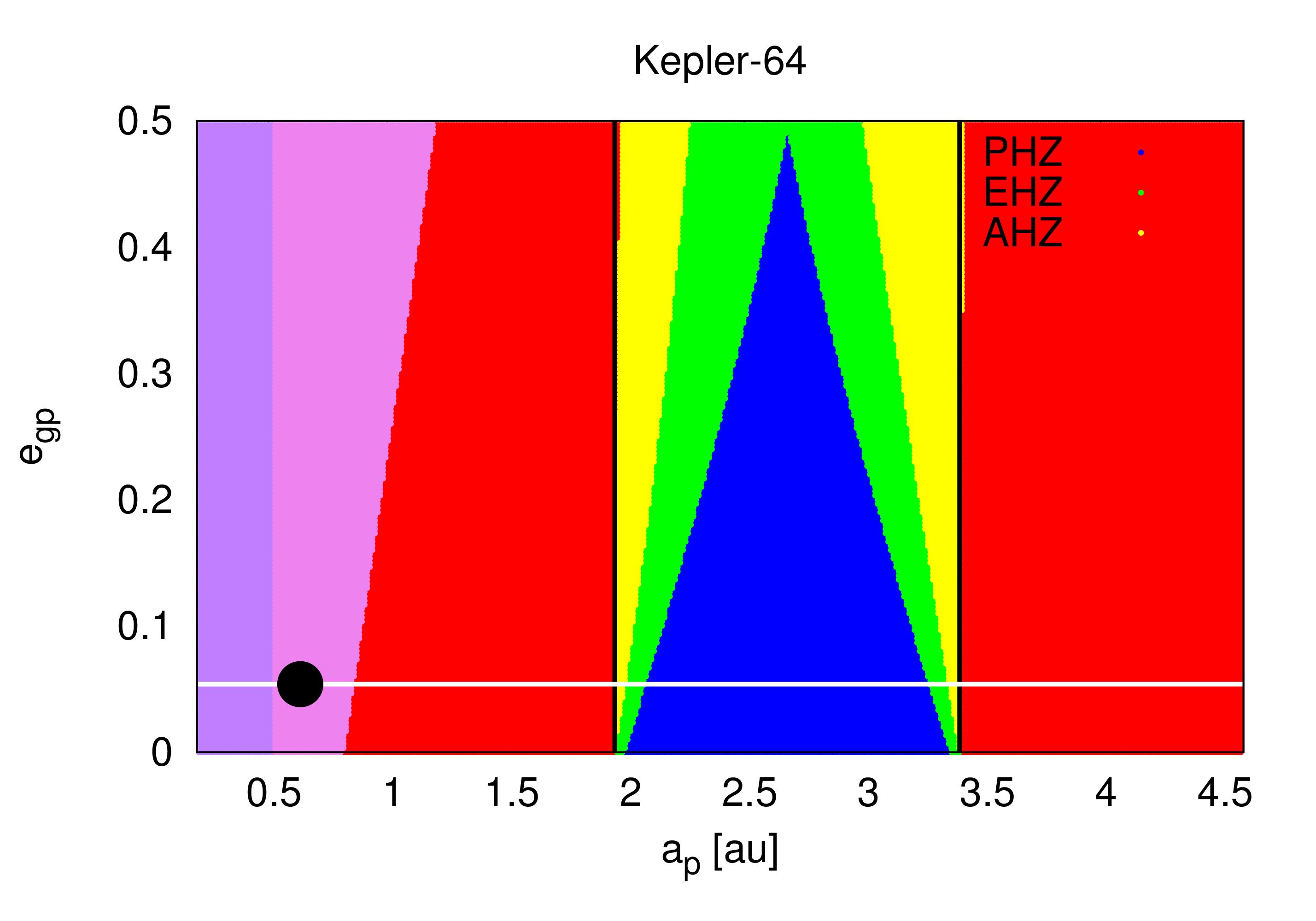}
\includegraphics[width=8cm]{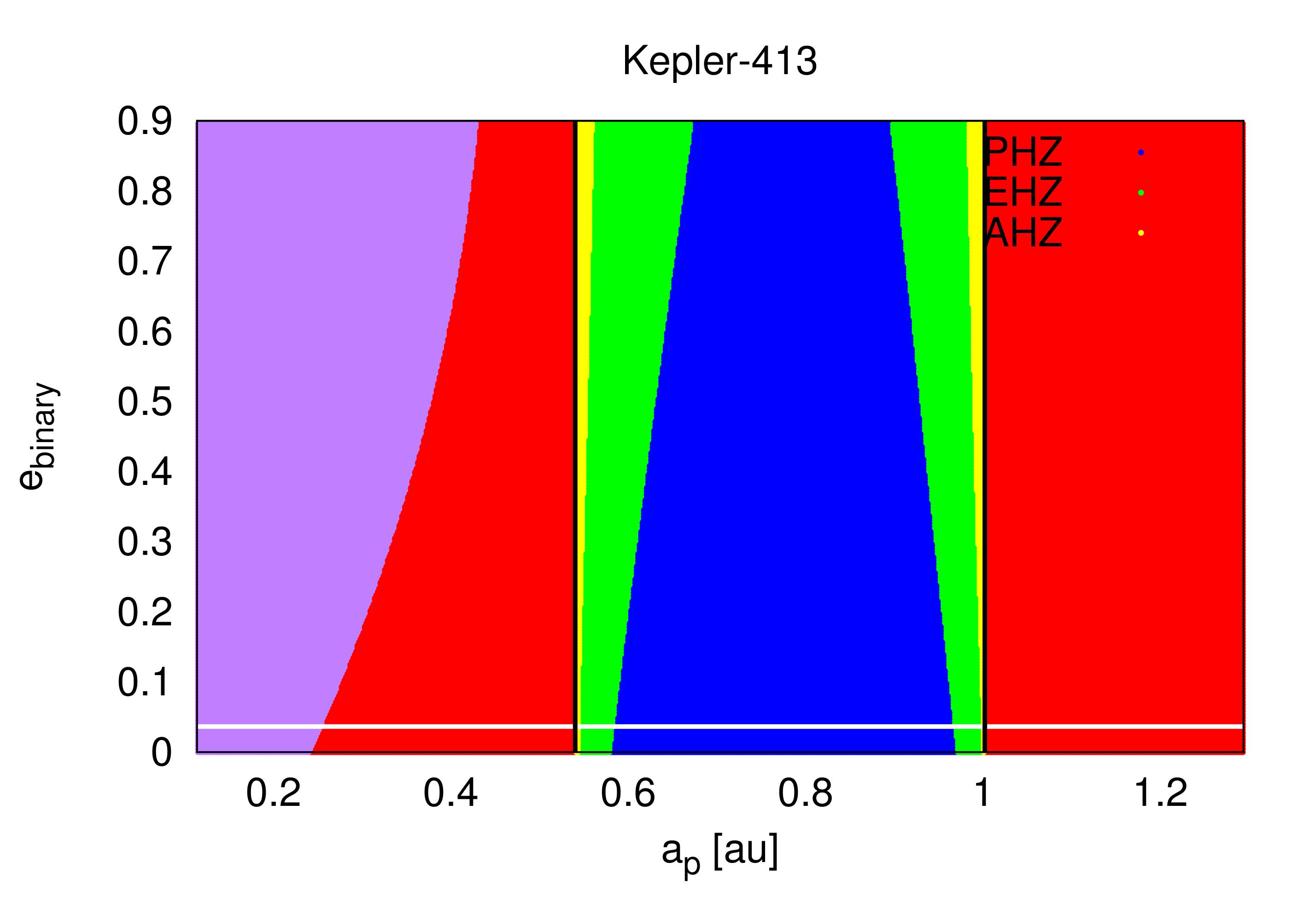}
\includegraphics[width=8cm]{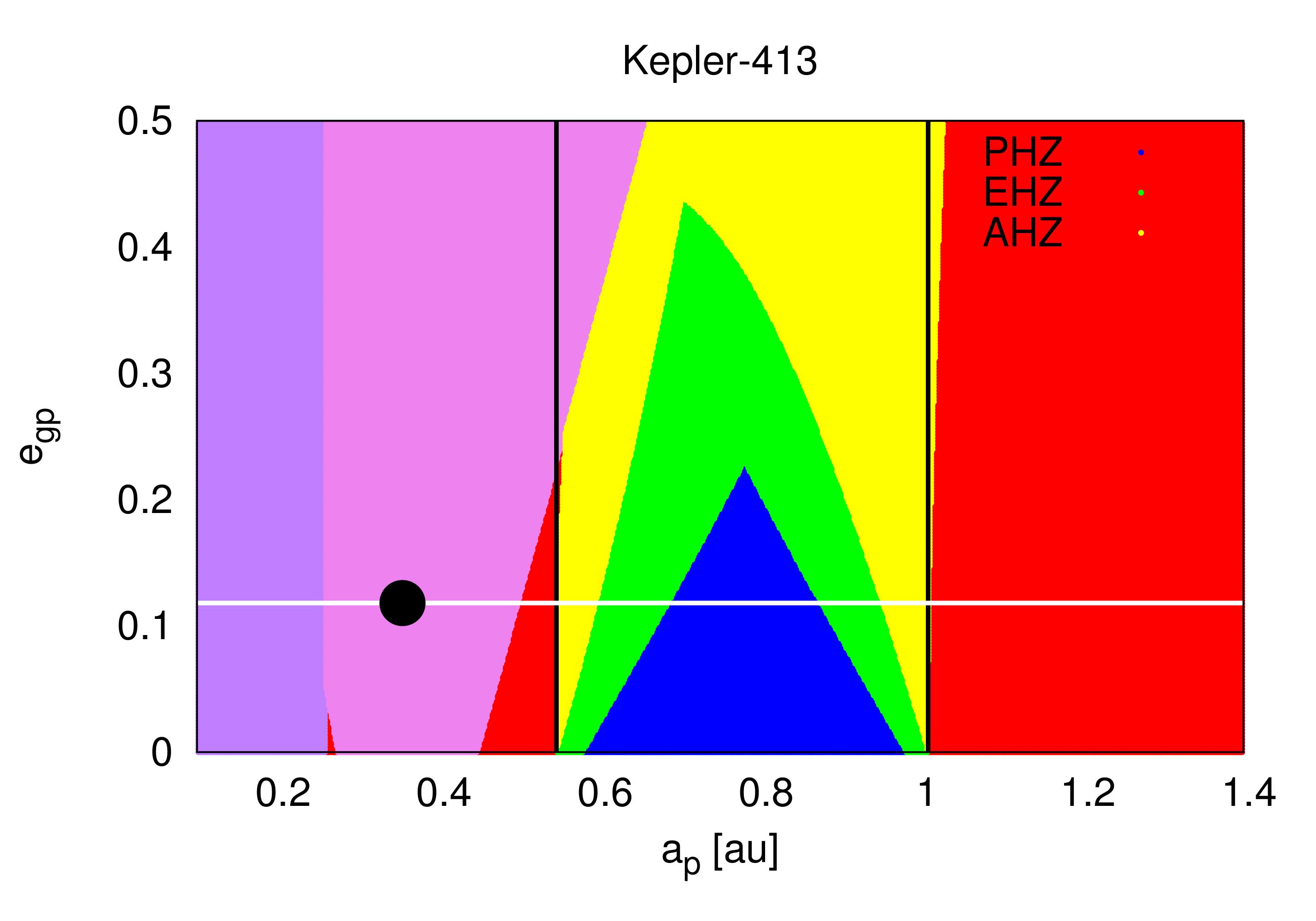}
\end{center}
\caption{Same as Figure \ref{fig:3} for Kepler-38, Kepler-64 and Kepler-413.}\label{fig:4}
\end{figure}

\begin{figure}[h!]
\begin{center}
\includegraphics[width=8cm]{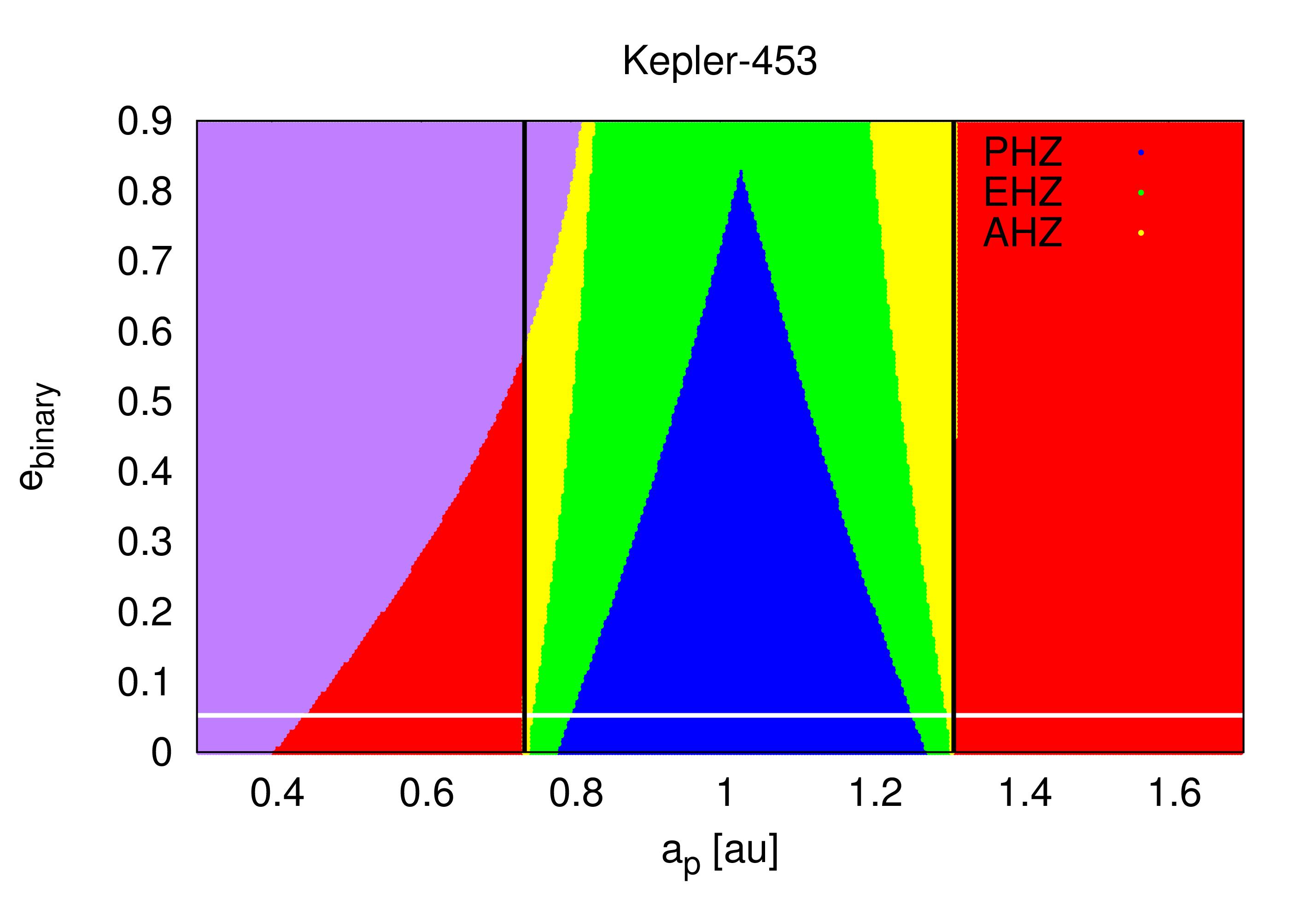}
\includegraphics[width=8cm]{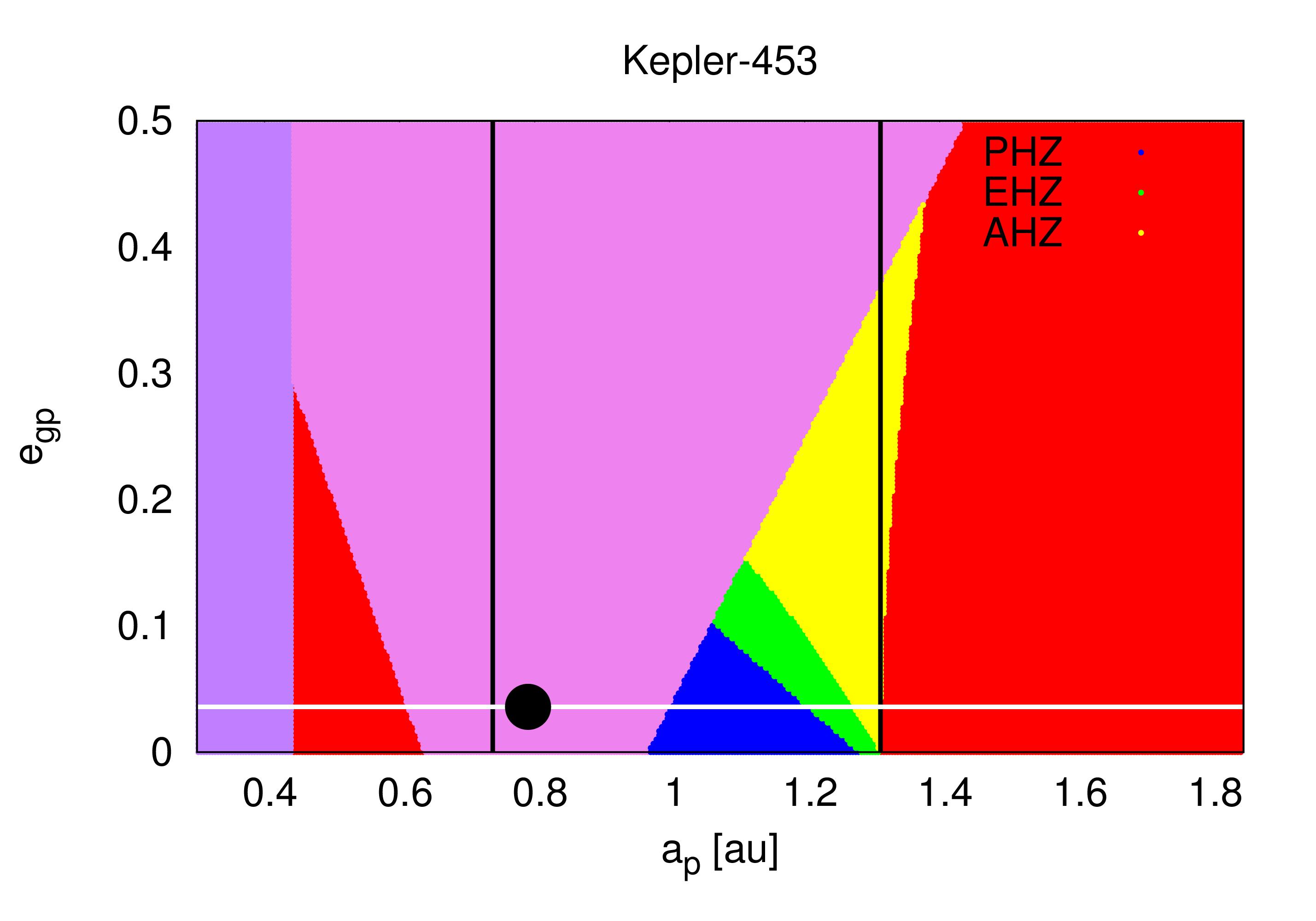}
\includegraphics[width=8cm]{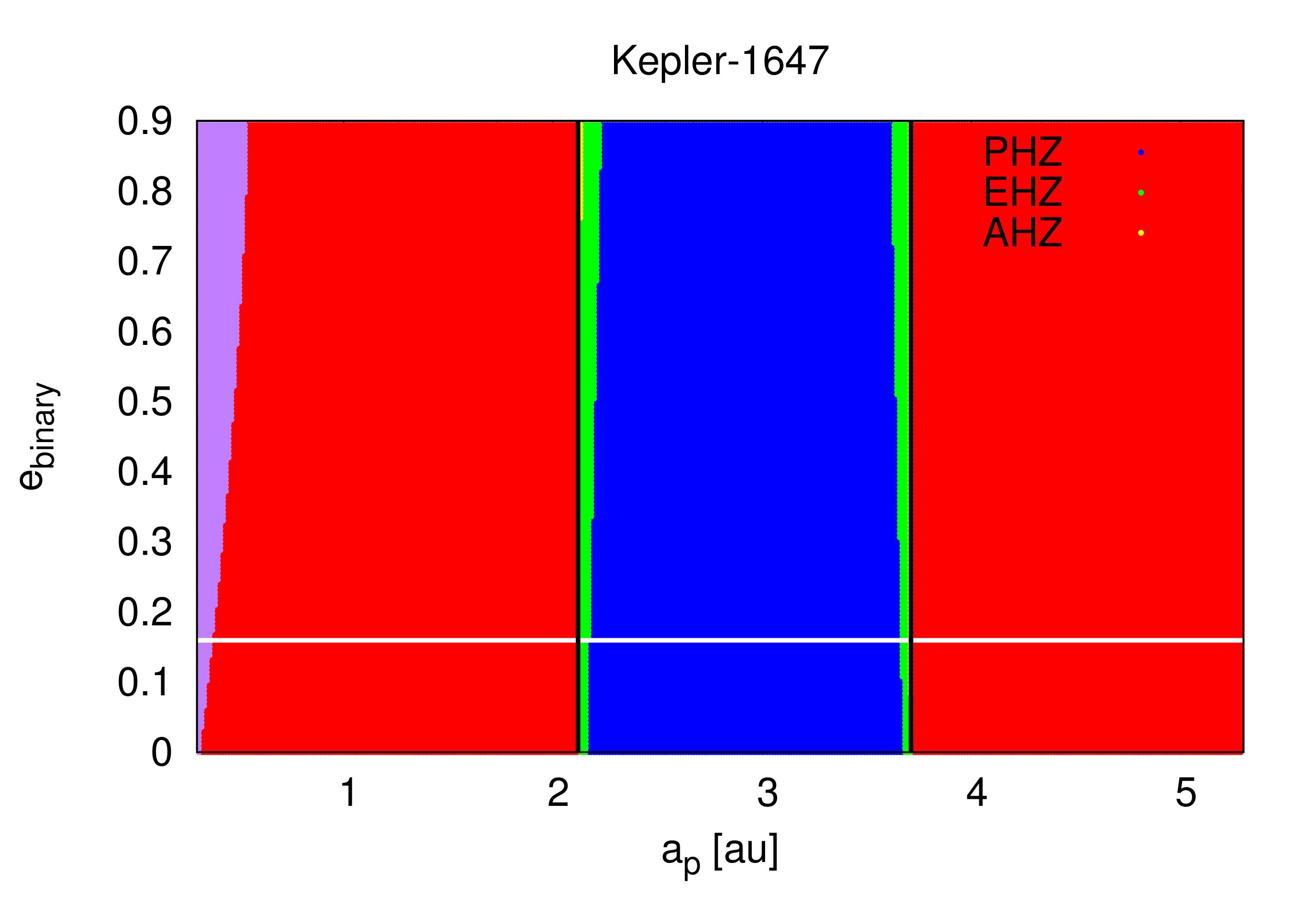}
\includegraphics[width=8cm]{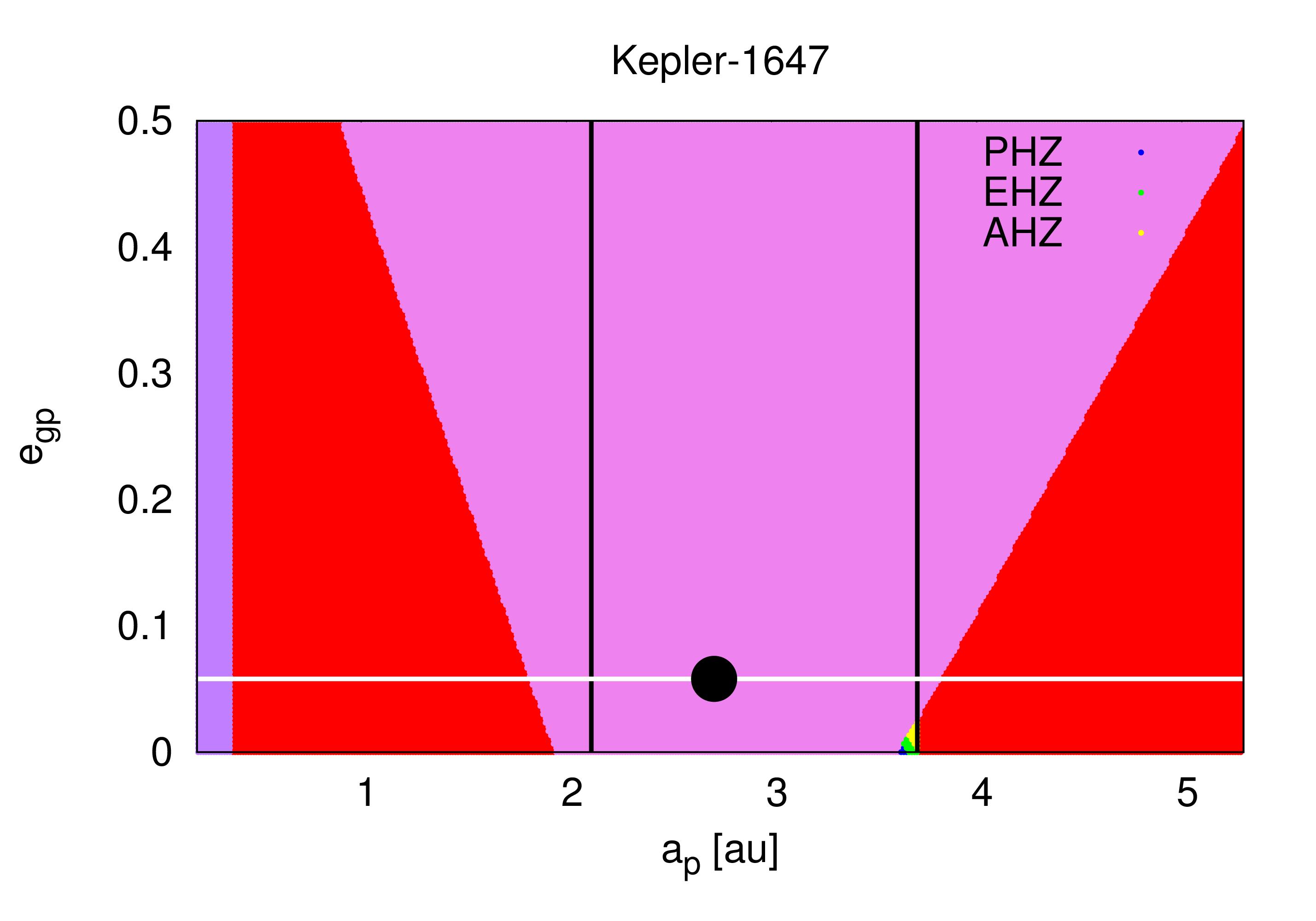}
\includegraphics[width=8cm]{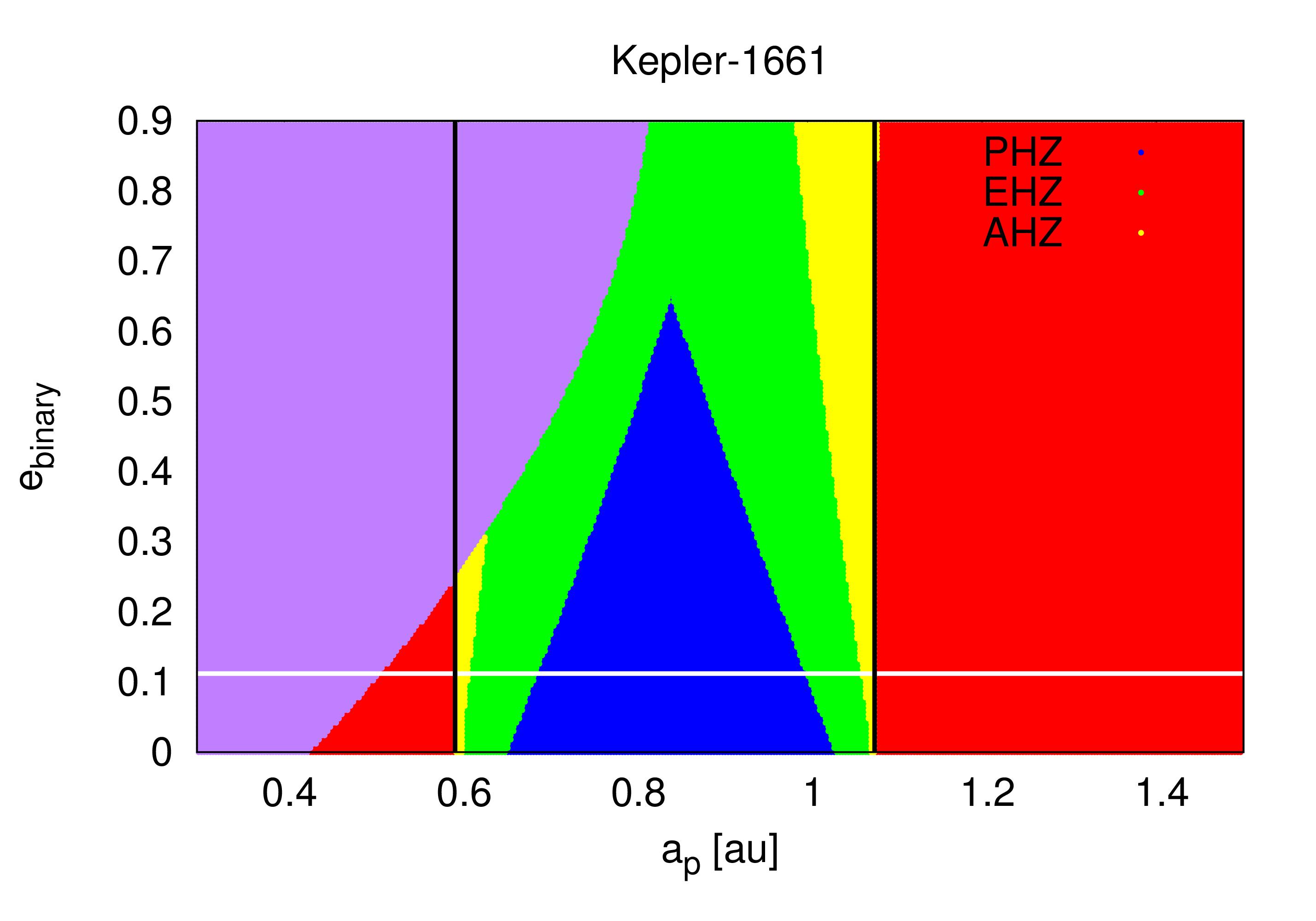}
\includegraphics[width=8cm]{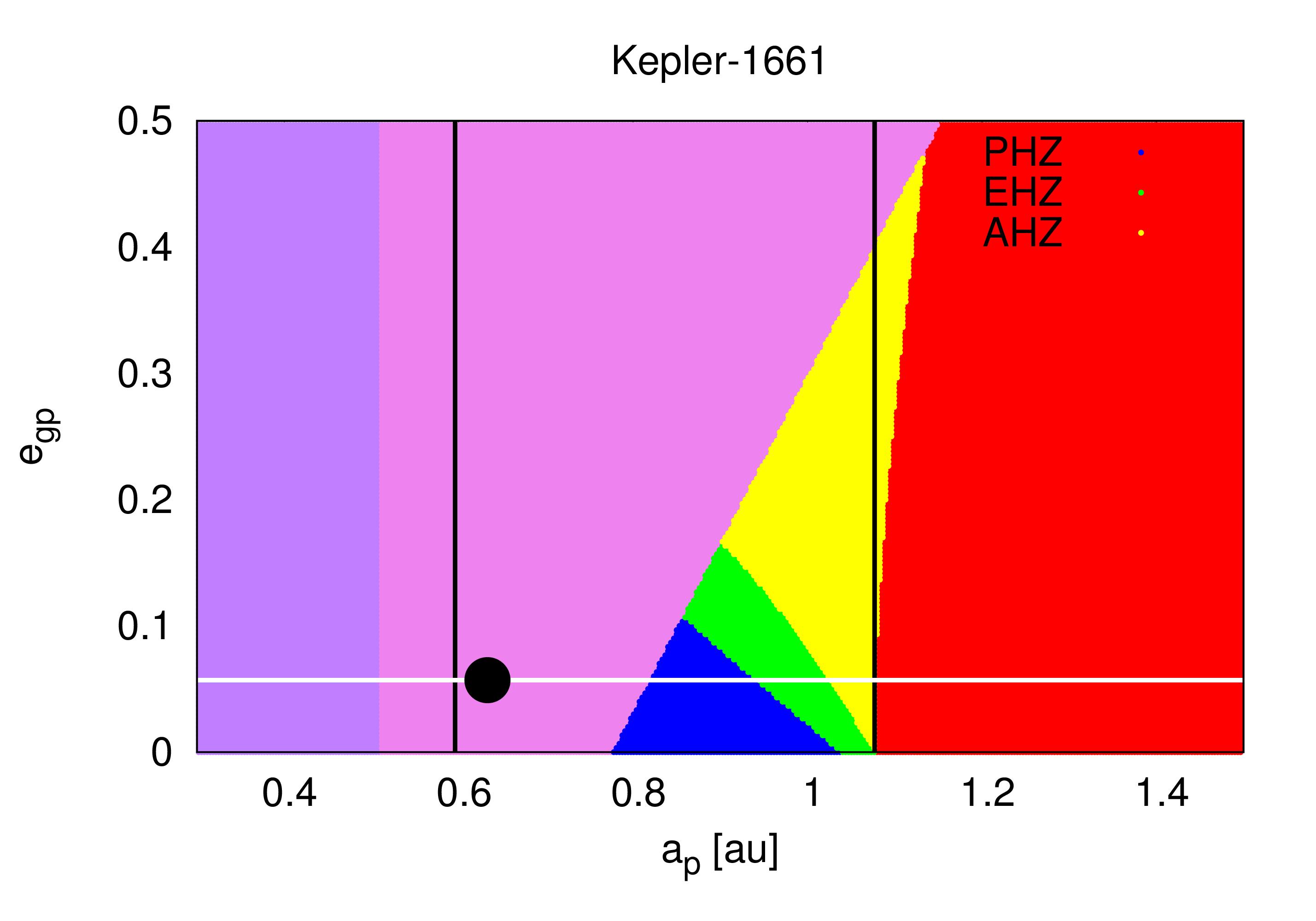}
\end{center}
\caption{Same as Figure \ref{fig:3} for Kepler-453, Kepler-1647 and Kepler-1661.}\label{fig:5}
\end{figure}

We now proceed to calculating habitable zones for the aforementioned Kepler systems. In a first step, we ignore the presence of residing giant planets in the respective systems.  Considering a potential terrestrial planet orbiting the stellar binary, we can determine the borders of the classical and dynamically informed habitable zones for all the Kepler systems under investigation. As we can see from the left column plots of Figures \ref{fig:3}, \ref{fig:4} and \ref{fig:5}, all systems but one have well defined habitable zones and would be capable of hosting a broad range of potentially habitable worlds. The only exception to that is Kepler-16, where more than half of the habitable zone is truncated due to dynamical instability arising from the gravitational perturbations of the stellar binary.  A terrestrial planet could only survive near the outer border of the habitable zone with the additional requirement that it does not have a low climate inertia since the PHZ has been completely eliminated.  

When we add the giant planet in our model (right column plots of Figures \ref{fig:3}, \ref{fig:4} and \ref{fig:5}), we notice that in Kepler-16 and Kepler-1647 the added gravitational effect of the giant planet renders the entire habitable zone dynamically unstable. Regarding Kepler-16, this is in agreement with other studies such as \citep{2012ApJ...750...14Q,2013ApJ...767L..38L}. Note that these results do not exclude the presence of potentially habitable moons around Kepler-16b. In the Kepler-1647 system, the giant planet is located near the middle of the habitable zone.
Although Kepler-453b and Kepler-1661b are located inside the classical habitable zones like the other two giant planets mentioned above, they allow a terrestrial planet to exist near the outer border of the habitable zone and
they may even allow that planet to have a partial PHZ.  This is because 
of the relatively small masses ($\sim$Neptune mass) and eccentricities of those planets. In contrast, Kepler-1647b is 
$1.5 M_J$ and resides near the center of the habitable zone. 

Regarding the remaining systems, the entire classical habitable zone is essentially dynamically stable.
The difference between dynamically informed habitable zones with and without the giant planet perturbers shows, however, that the influence of giant planets goes beyond dynamical instability. 
In Kepler-34, potentially habitable worlds with low climate inertia could only remain habitable in a tiny region centered around 2.17 au. This can be seen from a comparison between the extent of the PHZ and the black vertical lines in the middle row right panel of Figure \ref{fig:3}. Relatively dry planets with a low concentration of greenhouse gases would fall into this category. On the other hand, Kepler-35, Kepler-38 and Kepler-64 could host potentially habitable worlds with low climate inertia over a significant fraction (more than $75\%$) of their classical habitable zone (Figures \ref{fig:4} and \ref{fig:5}). Finally, the extent of the PHZ of Kepler-413 is around $40\%$ of its classical habitable zone.
While not prohibitive, the presence of giant planets in Kepler-34, Kepler-413, Kepler-453 and Kepler-1661 requires additional terrestrial worlds to buffer significant variations in irradiation in order to remain habitable.  

 For comparison, some results from \cite{2013ApJ...777..166H,2014ApJ...780...14C,2015ApJ...798..101C,2019ApJ...873..113W} are also provided.  Our results seem to be in good agreement with those of \cite{2014ApJ...780...14C,2015ApJ...798..101C,2019ApJ...873..113W}, with our classical habitable zone being a bit wider. In order to compare with \cite{2013ApJ...777..166H}, we use \cite{2013ApJ...770...82K} to calculate our classical habitable zone as that specific climate model was used by \cite{2013ApJ...777..166H} for their habitable zone calculations.  In this case, however, it is more difficult to make a direct comparison as  \cite{2013ApJ...777..166H} used time variable habitable zones centered at the primary star. Habitable zones borders for the Kepler systems under investigation are presented in Table \ref{tab3}.

\section{Discussion and summary}
\label{sec:discussion}
We present an analytical approach to determine dynamically informed habitable zones in binary star systems with a circumbinary giant planet. The method takes into consideration the orbital evolution of the giant and terrestrial planet as well as different responses of planetary climates to variations in the quantity and spectral energy distribution of incoming radiation. It does not apply, however, during the planet formation stage, when we can have planets migrating due to interactions with the protoplanetary disk or planet-planet scattering events during late stage formation. In addition, we do not consider systems where there is significant tidal interaction between the two stars  which may lead to changes in the orbit and rotation rates of the stars, as well as to changes in the emission of XUV radiation that can affect the atmosphere of a planet within the habitable zone \citep[e.g.][]{2014A&A...570A..50S,2019A&A...626A..22J}.
  
As the method mainly relies on analytical equations, it can provide a quick assessment of the capability of terrestrial circumbinary planets in complex dynamical environments to retain liquid water on their surface.  The construction and comparison of dynamically informed habitable zones, i.e. the PHZ, the EHZ and the AHZ allows us to better understand where potentially habitable worlds with different climate characteristics can exist in binary star systems. The method presented here is very versatile as it has been constructed in such a way that it does not depend on the dynamical model and the insolation limits.


In this work, we investigated the effects of stellar binarity and circumbinary giant planets on the habitable zones of nine systems observed by the Kepler mission. We confirm earlier studies that suggest Kepler-16 is not suitable for hosting a terrestrial planet within its classical habitable zone. The situation is similar for Kepler-1647. In contrast, Kepler-34, Kepler-35, Kepler-38, Kepler-64 and Kepler-413 seemed more promising with Kepler-38 being the best candidate in this respect. Kepler-453 and Kepler-1661 stand between the previous two categories of systems. We find that nearly equal binary mass ratios and small eccentricities of the perturbing bodies provide favorable, from the orbital evolution point of view, conditions for an Earth-like planet to exist in the habitable zone. We show, furthermore, that the presence of a giant planet can have a significant effect on the potential habitability of terrestrial worlds in the same system. We, thus, recommend gravitational perturbations of known giant planets to be taken into account in future studies regarding habitability in binary star systems.


\section*{appendix}
\subsection*{Planetary eccentricity equations}

We follow \cite{2015ApJ...802...94G} to calculate the maximum orbital eccentricity $e_p^{max}$ and average squared eccentricity $\langle e^2_p\rangle$ for a circumbinary planet: 
\begin{equation}
\label{app1}
e^{max}_{p}=\frac{m_{1}m_{2}}{(m_{1}+m_{2})^{\frac{4}{3}}M^{\frac{2}{3}}}\frac{1}{X^{\frac{4}{3}}}\left[\frac{3}{2}+
\frac{17}{2}e^2_b+\frac{1}{X}\left(3+19e_b+\frac{21}{8}e^2_b-\frac{3}{2}e^3_b\right)\right]+2\left|\frac{K_2}{K_1-K_3}\right|
\end{equation}
and
\begin{eqnarray}
\label{app2}
\langle e^2_p\rangle&=&\frac{m^2_{1}m^2_{2}}{(m_{1}+m_{2})^{\frac{8}{3}}M^{\frac{4}{3}}}\frac{1}{X^{\frac{8}{3}}}
\bigg[\frac{9}{8}+\frac{27}{8}e^2_b+\frac{887}{64}e^4_b-\frac{975}{64}\frac{1}{X}e^4_b\sqrt{1-e^2_b}+ \nonumber\\
& & +\frac{1}{X^2}\Big(\frac{225}{64}+\frac{6619}{64}e^2_b-\frac{26309}{512}e^4_b-\frac{393}{64}e^6_b\Big)\bigg]+2\Big(\frac{K_{2}}{K_1-K_3}\Big)^2,
\end{eqnarray}
where 
\begin{eqnarray}
M&=&m_1+m_2+m_p\\
X&=&\sqrt{\frac{m_1+m_2}{M}}\left(\frac{a_p}{a_b}\right)^{\frac{3}{2}} \\
K_1 & = & \frac{3}{8}\frac{\sqrt{\mathcal{G}M}m_1m_2a^2_{b}}{(m_1+m_2)^2a^{\frac{7}{2}}_{p}}(2+3e^2_{b})\\
K_2 & = &\frac{15}{64}\frac{\sqrt{\mathcal{G}M}m_1m_2(m_1-m_2)a^3_{b}}{(m_1+m_2)^3a^{\frac{9}{2}}_{p}}e_{b}(4+3e^2_{b})\\
K_3 & = & \frac{3}{4}\frac{\sqrt{\mathcal{G}}m_pa^{\frac{3}{2}}_{b}\sqrt{1-e^2_{b}}}{(m_1+m_2)^{\frac{1}{2}}a^3_{p}},
\end{eqnarray}
and $\mathcal{G}$ is the gravitational constant. 

Equations (\ref{app1}) (\ref{app2}) were also used when we added the giant planet to our model and the orbit of the giant planet was interior to that of the terrestrial planet. In order to use the above equations in that respect, we replace $m_1$ with $m_b=m_1+m_2$, $m_2$ with $m_{gp}$, $a_b$ with $a_{gp}$, $e_b$ with $e_{gp}$ and $M=m_b+m_{gp}+m_p$.

When the orbit of the giant planet was exterior to that of the terrestrial planet, we used the below equations from \cite{2012ApJ...752...74E,2003MNRAS.345..340G,2005MNRAS.362..748G} (with the notation of this work):

\begin{eqnarray}
e^{max}_p&=&\frac{m_{gp}}{MX^{5/3}}\left[\frac{15}{64}\frac{m_1-m_p}{(m_1+m_p)^{2/3} M^{1/3}}\frac{(4+11 e_{gp}^2)}{(1-e_{gp}^2)^{5/2}}+\frac{11}{4}\frac{1}{X^{1/3}}\frac{(1+e_{gp})^3}{(1-e_{gp}^2)^3}\right.+\nonumber\\
&&\left.+\frac{3}{4}\frac{1}{X^{4/3}}\frac{(1+e_{gp})^4(6+11 e_{gp})}{(1-e_{gp}^2)^{9/2}}\right]+2\frac{C}{B-A}, 
\end{eqnarray}

and

\begin{eqnarray}
\langle e_p^2 \rangle & = &
\frac{m_{gp}^{2}}{M^{2}}\frac{1}{X^{4}(1-e_{gp}^{2})^{\frac{9}{2}}}\left\{\frac{43}{8}+\frac{129}{8}e_{gp}^{2}+\frac{129}{64}e_{gp}^{4}+\frac{1}{(1-e_{gp}^{2})^{\frac{3}{2}}}\left(\frac{43}{8}+\frac{645}{16}e_{gp}^{2}+\frac{1935}{64}e_{gp}^{4}+\right. \right.\nonumber\\
& &\left.+\frac{215}{128}e_{gp}^{6}\right)+\frac{1}{X^{2}(1-e_{gp}^{2})^{3}}\left[\frac{365}{18}+\frac{44327}{144}e_{gp}^{2}+\frac{119435}{192}e_{gp}^{4}+\frac{256105}{1152}e_{gp}^{6}+\frac{68335}{9216}e_{gp}^{8}+\right.\nonumber\\
& & \left.+\frac{1}{(1-e_{gp}^{2})^{\frac{3}{2}}}\left(\frac{365}{18}+\frac{7683}{16}e_{gp}^{2}+\frac{28231}{16}e_{gp}^{4}+\frac{295715}{192}e_{gp}^{6}+\frac{2415}{8}e_{gp}^{8}+\frac{12901}{2048}e_{gp}^{10}\right)\right]+\nonumber\\
& & +\frac{1}{X(1-e_{gp}^{2})^{\frac{3}{2}}}\left[\frac{61}{3}+\frac{305}{2}e_{gp}^{2}+\frac{915}{8}e_{gp}^{4}+\frac{305}{48}e_{gp}^{6}+\frac{1}{(1-e_{gp}^{2})^{\frac{3}{2}}}\left(\frac{61}{3}+\frac{854}{3}e_{gp}^{2}+\right.\right.\nonumber\\
& & \left.\left.+\frac{2135}{4}e_{gp}^{4}+\frac{2135}{12}e_{gp}^{6}+\frac{2135}{384}e_{gp}^{8}\right)\right]+m_{*}^{2}X^{\frac{2}{3}}(1-e_{gp}^{2})\left[\frac{225}{256}+\frac{3375}{1024}e_{gp}^{2}+\frac{7625}{2048}e_{gp}^{4}+\right.\nonumber\\
& & +\frac{29225}{8192}e_{gp}^{6}+\frac{48425}{16384}e_{gp}^{8}+\frac{825}{2048}e_{gp}^{10}+\frac{1}{(1-e_{gp}^{2})^{\frac{3}{2}}}\left(\frac{225}{256}+\frac{2925}{1024}e_{gp}^{2}+\frac{775}{256}e_{gp}^{4}+\right.\nonumber\\
& &\left.\left.+\frac{2225}{8192}e_{gp}^{6}+\frac{25}{512}e_{gp}^{8}\right)\right]+m_{*}^{2}\frac{1}{X^{\frac{4}{3}}(1-e_{gp}^{2})^{2}}\left[\frac{8361}{4096}+\frac{125415}{8192}e_{gp}^{2}+\frac{376245}{32768}e_{gp}^{4}+\right.\nonumber\\
& & +\frac{41805}{65536}e_{gp}^{6}+\frac{1}{(1-e_{gp}^{2})^{\frac{3}{2}}}\left(\frac{8361}{4096}+\frac{58527}{2048}e_{gp}^{2}+\frac{877905}{16384}e_{gp}^{4}+\frac{292635}{16384}e_{gp}^{6}\right.\nonumber\\
& & \left.\left.\left.+\frac{292635}{524288}e_{gp}^{8}\right)\right]\right\}+2(\frac{C}{B-A})^{2},
\end{eqnarray}
where
\begin{eqnarray}
M&=&m_b+m_p+m_{gp}\\
m_*&=&\frac{m_p-m_b}{(m_b+m_p)^{2/3}M^{1/3}}\\
X&=&\sqrt{\frac{m_b+m_p}{M}}\left(\frac{a_{gp}}{a_p}\right)^{\frac{3}{2}} \\
C &=& \frac{5}{4}\frac{(m_b-m_p)e_{gp}a_p}{(m_b+m_p)(1-e^2_{gp})^{5/2}a_{gp}}\\
B &=& \frac{1}{(1-e^2_{gp})^{3/2}}\\
A &=& \frac{m_bm_pM^{1/2}}{m_{gp}(m_b+m_p)^{3/2}}\frac{1}{(1-e_{gp})^2}\left(\frac{a_p}{a_{gp}}\right)^{1/2}.
\end{eqnarray}

The above equations can be used for nearly coplanar systems and systems that are not close to mean motion resonances. 
Also, the equations become less reliable as the maximum planetary eccentricity gets larger than 0.2-0.25.

\subsection{Figures}

\section*{Author Contributions}

Conceptualization, N.G. and S.E; methodology, N.G. AND S.E.; software, N.G.; validation, N.G., S.E. and I.D.-D.; investigation, N.G. and S.E.; resources, N.G. AND S.E.; data curation, N.G.; writing—original draft
preparation-revised version, N.G. AND S.E.; All authors have read and agreed to the published version of the manuscript.

\section*{Acknowledgments}
This research has made use of the NASA Exoplanet Archive, which is operated by the California Institute of Technology, under contract with the National Aeronautics and Space Administration under the Exoplanet Exploration Program.
SE acknowledges support from the DIRAC Institute in the Department of Astronomy at the University of Washington. The DIRAC Institute is supported through generous gifts from the Charles and Lisa Simonyi Fund for Arts and Sciences, and the Washington Research Foundation.
The results reported herein benefited from collaborations and/or information exchange within NASA's Nexus for Exoplanet System Science (NExSS) research coordination network sponsored by NASA's Science Mission Directorate.

\bibliographystyle{frontiersinSCNS_ENG_HUMS} 
\bibliography{ref}



\end{document}